\documentclass[aps,amsfonts,pra,twocolumn,showpacs]{revtex4-2}
\usepackage{epsfig,amsmath,amssymb,bm,epsf,graphicx,psfrag,float}

\usepackage[all]{xy}
\usepackage{xcolor}

\usepackage{hyperref}
\usepackage{qcircuit}

\def\ketbra#1{\vert#1\rangle\langle#1\vert}

\newcommand{\tprod}{\mathop{\overleftarrow{\prod}}}
\def\ketbra#1{\Big\vert#1\Big\rangle\Big\langle#1\Big\vert}

\begin{document}

\title{Measurement-Based Preparation of Higher-Dimensional AKLT States and Their Quantum Computational Power}

\author{Wenhan Guo}
\affiliation{C. N. Yang Institute for Theoretical Physics, State University of New York at Stony Brook, Stony Brook, NY 11794-3840, USA}
\affiliation{Department of Physics and Astronomy, State University of New York at Stony Brook, Stony Brook, NY 11794-3800, USA}
\author{Mikhail Litvinov}
\affiliation{C. N. Yang Institute for Theoretical Physics, State University of New York at Stony Brook, Stony Brook, NY 11794-3840, USA}
\affiliation{Department of Physics and Astronomy, State University of New York at Stony Brook, Stony Brook, NY 11794-3800, USA}
\author{Tzu-Chieh Wei}
\affiliation{C. N. Yang Institute for Theoretical Physics, State University of New York at Stony Brook, Stony Brook, NY 11794-3840, USA}
\affiliation{Department of Physics and Astronomy, State University of New York at Stony Brook, Stony Brook, NY 11794-3800, USA}

\author{Abid Khan}
\affiliation{Department of Physics, University of Illinois Urbana-Champaign, Urbana, IL, United States 61801}
\author{Kevin C. Smith}
\affiliation{IBM Quantum, IBM Research Cambridge, Cambridge, MA 02142, USA}

\date{\today}

\begin{abstract}
We investigate a constant-time, fusion measurement-based scheme to create AKLT states beyond one dimension. We show that it is possible to prepare such states on a given graph up to random spin-1 `decorations', each corresponding to a probabilistic insertion of a vertex along an edge.  In investigating their utility in measurement-based quantum computation, we demonstrate that any such randomly decorated AKLT state possesses at least the same computational power as non-random ones, such as those on trivalent planar lattices.
For AKLT states on Bethe lattices and their decorated versions we show that there exists a deterministic, constant-time scheme for their preparation.
In addition to randomly decorated AKLT states, we also consider random-bond AKLT states, whose construction involves any of the canonical Bell states in the bond degrees of freedom instead of just the singlet in the original construction.  Such states naturally emerge upon measuring all the decorative spin-1 sites in the randomly decorated AKLT states. We show that those random-bond AKLT states on trivalent lattices can be converted to encoded random graph states after acting with the same POVM on all sites. We also argue that these random-bond AKLT states possess similar quantum computational power as the original singlet-bond AKLT states via the percolation perspective.
\end{abstract}

 \maketitle

 \section{Introduction}

The Affleck–Kennedy–Lieb–Tasaki (AKLT) models~\cite{AKLT,AKLT2} have provided useful insights in various contexts~\cite {wei2022some}. For example, the spin-1 model on a one-dimensional chain provides confirmation
of Haldane's conjecture~\cite{Haldane83,Haldane83b} via an exact model construction that gives a provable excitation gap above the constructed ground state.  The spin-1 AKLT state thus belongs to the Haldane phase and is recognized to possess nontrivial symmetry-protected topological (SPT) order~\cite{Gu,Chen,Pollmann,Chen2013,ElseNayak}. Likewise, the 2D AKLT states on the honeycomb and square lattices have, respectively, weak and strong SPT order, e.g., characterized by the strange correlator~\cite{YouBiRasmussenSlagleXu2014,WierschemBeach2016}. 

Due to the valence-bond-solid (VBS) construction using spin singlets,  AKLT states are generally expected to be magnetically disordered, as shown in one dimension~\cite{AKLT} and two dimensions~\cite{parameswaran2009order}, but they can also exhibit N\'eel order, e.g., on Bethe lattices with high coordination numbers~\cite{AKLT2,Pomata} and the three-dimensional cubic lattice~\cite{parameswaran2009order}. They can even form ordered quantum vector spin glass states on random graphs of fixed connectivity~\cite{laumann2010aklt}. AKLT models can be generalized under deformation, enabling study of the quantum phase transition between the VBS and N\'eel phases using classical vertex models ~\cite{NiggemannKlumperZittartz1997,NiggemannKlumperZittartz2000} and tensor networks~\cite{huang2015emergence,PomataHuangWei18} on various lattices, including the uncovering of an XY-like phase in the deformed model on the square lattice.

The existence of the gap in some 2D and 3D AKLT models has recently been established rigorously~\cite{abdul-rahman2020analytic,pomata2019aklt,pomata2020demonstrating,lemm2020existence,guo2021nonzero}. Such proofs solidify the robustness of the magnetic disordering in these models. Moreover, various AKLT states were previously shown to enable universal measurement-based quantum computation (MBQC), including states in 2D trivalent lattices, such as the hexagonal lattice, the square lattice, and various decorated lattices~\cite{WeiAffleckRaussendorf11,Miyake11,Wei13,WeiPoyaRaussendorf,WeiRaussendorf15}.  The AKLT construction has also been used to produce quantum many-body scar states~\cite{moudgalya2018entanglement,lin2019exact,shiraishi2019connection,mark2020unified}. These exemplify the utility of AKLT models.

Given the significant interest in them, it is worth exploring the preparation of AKLT and related states. Recent progress in NISQ devices has sparked interest in this problem, and several recent works have studied it~\cite{malz2024preparation, murta2022preparing, smith2022deterministic}. 
Because it is a matrix product state (MPS) ~\cite{perez-garcia2007matrix}, the 1D AKLT state can be exactly prepared with a known quantum circuit ~\cite{schon2005sequential}. However, this scheme requires a time complexity linear in the system size; several works have since improved upon this result. Ref.~\cite{malz2024preparation} showed that faithfully preparing translation-invariant, short-range entangled MPS with non-zero correlation length (such as the 1D AKLT state) generically requires a unitary circuit of depth $\Omega(\log  N)$, where $N$ is the number of sites. While this bound appears to set strict limits on the efficiency with which the 1D AKLT state can be prepared, interestingly, this turns out not be the case. As shown by Ref.~\cite{smith2022deterministic}, it can be prepared deterministically in constant depth by leveraging midcircuit measurements and classical feedforward in addition to local unitary gates. Subsequently, Refs.~\cite {smith2024constant,sahay2025classifying,stephen2024preparingmatrixproductstates,zhang2024characterizing} extended this preparation technique to a broad class of MPS beyond the 1D AKLT state, demonstrating the promise of measurement-based schemes for preparing complex states in low depth.

In contrast to the 1D case, higher-dimensional AKLT states are evidently harder to prepare. Ref.~\cite{murta2022preparing} suggests that AKLT states in 2D can be prepared with a unitary circuit depth linear in the system size, but with a probability of success that is exponentially small.
Mirroring progress in 1D, it is reasonable to ask whether measurement-based techniques can aid in the preparation of higher-dimensional AKLT states. However, it is currently unclear whether the techniques applied in 1D can be extended to higher dimensions. The key issue intuitively lies in the existence of loops in two or higher dimensions, leading to uncorrectable defects that become ``trapped'' (and a non-deterministic protocol as a result). A similar feature occurs in 1D in the case of periodic boundary conditions ~\cite{smith2022deterministic,smith2024constant}.  

Here, we construct schemes to realize AKLT states in one, two, and possibly higher dimensions.  First, by extending the techniques used in 1D, we show how to {\it deterministically\/} prepare AKLT states and their deformed variants on Bethe lattices (which have no loops). Following this, we turn our attention to more general lattices in 2D. Instead of addressing the ``loop problem'' discussed above, we adopt a relaxed objective: rather than preparing exact AKLT states on a fixed graph, we allow for modifications of the underlying lattice. In particular, edges may be randomly decorated with one or more additional spin‑1 sites, corresponding to degree‑2 vertices that arise as post‑measurement defects. The resulting \emph{decorated} AKLT states are generated randomly but with a constant depth. Moreover, we will show that they have at least the same quantum computational universality as the original AKLT states on the ideal, decoration-free graph. 

Related to the decorated AKLT states, we also discuss the preparation of what we term \emph{random-bond} AKLT states. These are similar to the original AKLT states constructed from singlet bonds, but here the bonds between any pair of virtual qubits on neighboring sites can be any of the four Bell states (one singlet and three triplet entangled states). They appear naturally from one of our schemes when joining building blocks by measuring the neighboring dangling (virtual) qubits in the Bell basis. They can also be regarded as arising from measurements of a decorated AKLT state; upon measuring the spin-1 decorations in the virtual triplet Bell state basis (or equivalently in the basis $|+1\rangle \pm |-1\rangle$ \& $|0\rangle$, involving spin-1 states with $S_z=\pm1\, \&\, 0$), each is replaced with a random Pauli byproduct along that bond. For these states, we show that after applying the same POVM on all sites, as done in Ref.~\cite{WeiAffleckRaussendorf11}, the post-POVM state is an encoded (random) graph state. We then examine whether they can also be used as a universal resource for MBQC and argue that, as long as the original AKLT state on a trivalent lattice is universal for measurement-based quantum computation, the random-bond AKLT state on the same lattice will be universal as well. 

The remainder of the paper is structured as follows. In Sec.~\ref{sec:AKLT}, we review the original construction of AKLT with singlet bonds and generalize it to have bonds corresponding to arbitrary Bell states. 
In Sec.~\ref{sec:1D}, we discuss (deformed) 1D AKLT states and present an efficient scheme to prepare them in constant time.  
We also discuss another example in the case of an AKLT quasichain. In Sec.~\ref{sec:arbitrary}, we comment on the preparation of the 1D AKLT state with periodic boundary conditions, using this case as a preclude to the challenges loops pose for higher-dimensional AKLT states. 
In Sec.~\ref{sec:Blocks}, we present the ingredients for preparing higher-dimensional AKLT states: first, the unitary preparation of small elementary ``building block'' states and second, the technique for merging them into a larger AKLT state using fusion measurements.
In particular, we show that it is possible to deterministically prepare AKLT states (and, more generally, their deformed counterparts) on the Bethe lattices.  In Sec.~\ref{sec:Universality}, we discuss the quantum computational universality for two types of AKLT states: ones with random decoration and ones with random bonds. We narrow our focus on 2D lattices and show that, despite the
preparation of individual states not being deterministic,
 the ensemble of prepared states is universal.   In Sec.~\ref{sec:Conclude}, we make concluding remarks. In the Appendices, we supplement with additional materials.
\section{Affleck–Kennedy–Lieb–Tasaki (AKLT) states}
\label{sec:AKLT}
\vspace{-0.1cm}
\begin{figure*}
\centering
\vspace{-0.2cm}
\includegraphics[width=0.99\textwidth]{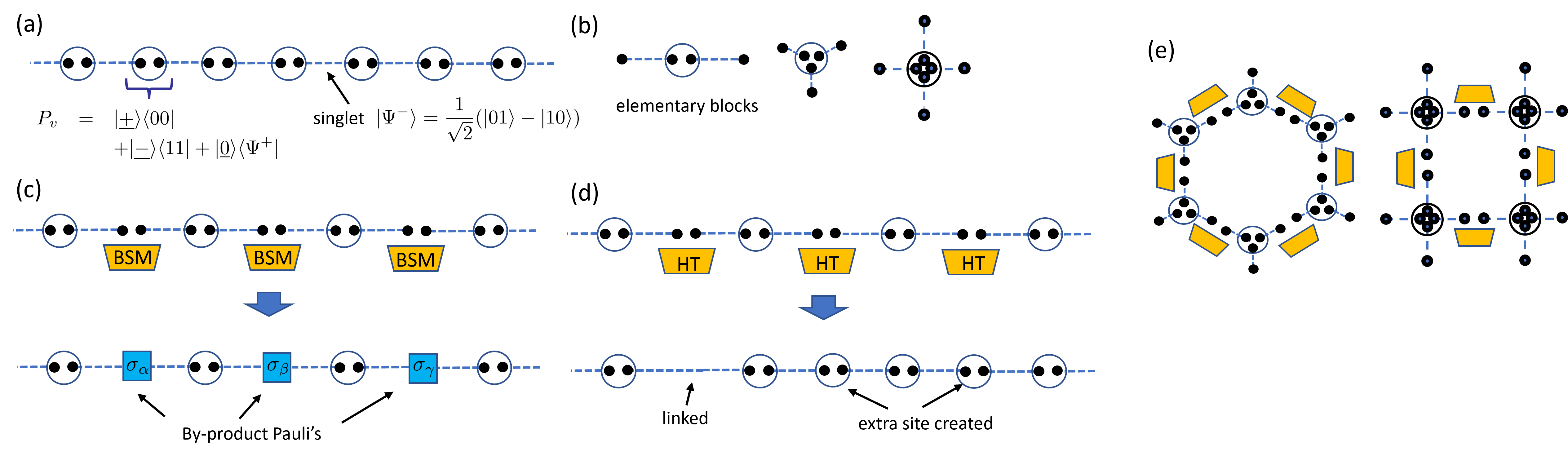}
\vspace{-0.35cm}
\caption{1D AKLT state and a scheme to create it by fusing elementary blocks. Each dot represents a qubit, and two dots connected by a dashed line is a two-qubit single state: $|\Psi^-\rangle=(|01\rangle-|10\rangle)/\sqrt{2}$. A circle represents a physical site, and its actual physical degree of freedom corresponds to the symmetric subspace of the qubits inside the circle. (a) Schematic definition of 1D spin-1 AKLT state. The blue circles here represent the mapping from virtual qubits to physical spins. (b) Several elementary blocks to be used in the fusion procedure, shown in (c), (d) \& (e), which uses Bell-state measurement (BSM) in (c) and Hadamard test (HT) in (d), indicated by yellow trapezoids, on two dangling qubits to fuse the two blocks into a larger part of the AKLT state.  The measurement in (e) can be BSM or HT. Illustration for creating (c) 1D 
and (e) 2D AKLT states.} 
\label{fig:AKLT1D}
\end{figure*}

\begin{figure*}[ht!]
\centering
\vspace{-0.2cm}
\includegraphics[width=0.99\textwidth]{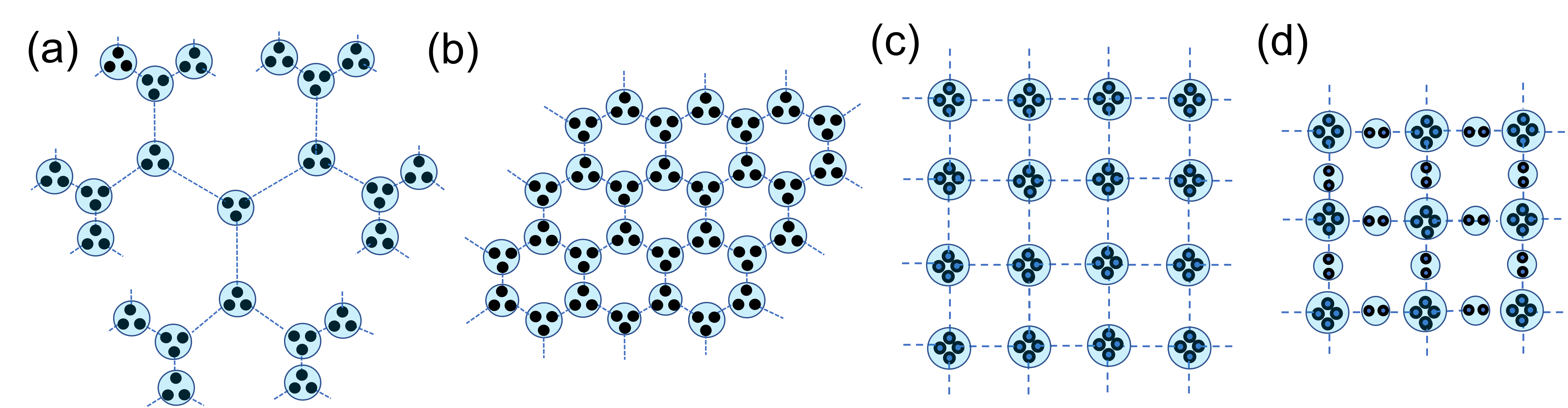}
\vspace{-0.1cm}
\caption{Illustration of 2D AKLT states.  (a) Bethe lattice with coordination number $z=3$,  (b) Hexagonal lattice, (c) square lattice, and (d) (uniformly) decorated square lattice. Each dot inside a circle represents a qubit or a spin-1/2 entity. A circle represents a physical site, which corresponds to spin magnitude $S=z_v/2$ (equivalently, the symmetric subspace of the qubits inside the circle), where $z_v$ is the number of dots inside the circle. } 
\label{fig:2DAKLT}
\end{figure*}

Let us begin with some definitions and background.
AKLT states can be defined on any graph. In Figs.~\ref{fig:AKLT1D} and~\ref{fig:2DAKLT}, we show examples of 1D and 2D AKLT states, as well as some building blocks. To define the AKLT wavefunction, we imagine that, first, each site $v$ consists of $z_v$ virtual qubits (spin-1/2's) with $z_v$ being the number of its neighbors, and one virtual qubit and its neighboring one along an edge $e$ (which shall be referred to as a bond) form a singlet $|\Psi^-\rangle_e=(|01\rangle-|10\rangle)/\sqrt{2}$  Then, there is a mathematical onsite projection that takes $z_v$ virtual qubits to their symmetric subspace, corresponding to a physical spin of total $S=z_v/2$. The basis states in the symmetric subspace of $z_v=n$ qubits can be constructed as 
\begin{equation}|D(n,k)\rangle=\sum_{\rm distinct \atop permutations} |\underbrace{0\dots0}_{n-k}\underbrace{1\dots1}_{k}\rangle/\sqrt{C^n_k},
\end{equation}
and they correspond to physical states $|S=n/2, m_s=(n-2k)/2\rangle$. We refer the readers to the original papers~\cite{AKLT,AKLT2} or the review chapter~\cite{wei2022some}.

For the 1D AKLT state, each  spin-1 site can be constructed by adding two spin-1/2's as
$|\underline{+}\rangle\equiv |S=1,S_z=+1\rangle=|00\rangle$, $|\underline{-}\rangle\equiv |S=1,S_z=-1\rangle=|11\rangle$, and 
$|\underline{0}\rangle\equiv |S=1,S_z=0\rangle=|\Psi^+\rangle$.
The 1D AKLT state is the (unique) ground state of the  Hamiltonian with periodic boundary condition~\cite{AKLT},
\begin{equation}
 H_{\rm AKLT}^{S=1}=\frac{1}{2}\sum_j\Big[\vec S_j\cdot \vec S_{j+1}+\frac{1}{3}(\vec S_j\cdot \vec S_{j+1})^2+S_{j+1}+\frac{2}{3}\Big].\end{equation}
 The 1D AKLT wavefunction can be expressed in terms of a matrix product state, and 2D or higher-dimensional AKLT states can be represented as a tensor-network or projected entangled-pair state; see Fig.~\ref{fig:2DAKLT} for examples of AKLT states on Bethe lattices, 2D lattices,  and lattices with decorations on edges.
Such a construction of the ground state and its parent Hamiltonian works on any graph; for a vertex $v$ that has $z$ neighbors, the local Hilbert space corresponds to that of a spin-$z/2$ entity, equivalent to the symmetric subspace of $z$ spin-1/2 particles or qubits, and the parent Hamiltonian consists of terms that are polynomials of $\vec{S}_i\cdot \vec{S}_j$, for $(i,j)$ being an edge. The projection to the symmetric subspace equivalent to spin-$z/2$ in terms of qubit degrees of freedom is 
\begin{equation}
\label{eq:Projection}
    P_v =\sum_{k=0}^{z} |D(z,k)\rangle\langle D(z,k)|.
\end{equation}
To describe the mapping between virtual qubits and physical spin, we can rewrite the above as
\begin{equation}
    \tilde{P}_v =\sum_{k=0}^{z} |S=z/2,S_z=(z-2k)/2\rangle\langle D(z,k)|.
\end{equation}

The AKLT states turn out to be related to cluster/graph states. 
It was first shown in Ref.~\cite{Chen2010quantum} that by an adaptive local measurement scheme, the one-dimensional spin-1 AKLT state can be converted to a one-dimensional cluster state (with variable lengths).
This was subsequently shown to be possible with a fixed POVM measurement~\cite{WeiAffleckRaussendorf11}, whose POVM consists of $\{F^\dagger_x F_x, F^\dagger_y F_y, F^\dagger_z F_z\}$, with $\sum_{\alpha} F_\alpha^\dagger  F_\alpha =I_{S=1}$ being the identity operator for the spin-1 degree of freedom, and the $F$'s being described by
 \begin{eqnarray}
&&{F}^{S=1}_{\alpha}=\sqrt{\frac{1}{2}}\left(\ketbra{S_\alpha=+1}_\alpha\right.\nonumber\\
&& \quad\quad+\left.\ketbra{S_\alpha=-1}_\alpha\right) \label{eq:POVMS1}\\ &&=\sqrt{\frac{1}{2}} \left( \ketbra{D(2,0)}_\alpha+\ketbra{D(2,2)}_\alpha\right), \nonumber
\end{eqnarray}
where we have used the subscript $\alpha$ to represent which quantization axis $\alpha=x,y,z$ is used to define the state $|D(n,k)\rangle$. The top line uses the spin-1 picture, and the second line uses the virtual two-qubit picture.
A similar POVM on spin-3/2 entities allows to show that the spin-3/2 AKLT states on trivalent lattices, such as the hexagonal one, can be converted to a random graph state, which can be subsequently further converted to a regular 2D cluster state~\cite{WeiAffleckRaussendorf11,Wei13}.  The $F$ operators used in the spin-3/2 POVM are described by
 \begin{eqnarray}
{F}^{S=3/2}_{\alpha}&=&\sqrt{\frac{2}{3}}\left(\ketbra{\frac{3}{2}}_\alpha+\ketbra{-\frac{3}{2}}_\alpha\right) \nonumber \\ & = & \sqrt{\frac{2}{3}} \sum_{k=0,3} \ketbra{D(3,k)}_\alpha. \label{eq:POVM}
\end{eqnarray}
One might naively expect that one could generalize this to any spin-$S$, but it turns out that the relation $\sum_{\alpha} F_\alpha^\dagger  F_\alpha = I_{S}$ does not hold beyond $S=3/2$. For the $S=2$ case, a simple resolution exists: one can add terms whose action preserves the logical subspace of the encoded graph state \cite{WeiRaussendorf15}. This approach was used to show that the spin‑2 AKLT state on the square lattice remains universal for MBQC.  We are not aware of any similar resolution beyond $S=2$.

It is worth mentioning that even though $\sum_{\alpha} F_\alpha^\dagger  F_\alpha$ is generally not proportional to the identity operator $I_{S}$ for higher spins $S$, it was shown~\cite{wei2012two} that the following state
\begin{equation}
|\Psi_G\rangle\equiv\mathop{\bigotimes}_v F_{\alpha_v,v}\mathop{\bigotimes}_v P_v\mathop{\bigotimes}_e|\Psi^-\rangle_e=\mathop{\bigotimes}_vF_{\alpha_v,v}\mathop{\bigotimes}_e|\Psi^-\rangle_e
\end{equation}
is 
still an encoded graph state for arbitrary labels $\{\alpha_v=x,y,{\rm or}\, z\}$'s that represent the POVM outcomes, where we have used the properties that $F_{\alpha_v,v}P_v=F_{\alpha_v,v}$, as $P_v$ is the projection to the symmetric subspace and $F_{\alpha_v,v}$ is a finer-grained projection in the symmetric subspace.

In anticipating the discussion on the random-bond AKLT states below, we note that instead of $|\Psi^-\rangle$, other Bell states $|\Psi^+\rangle=(|01\rangle+|10\rangle)/\sqrt{2}$ and $|\Phi^\pm\rangle=(|00\rangle\pm|11\rangle)/\sqrt{2}$ can also be used in the bond, which was explored earlier in Ref.~\cite{huang2015emergence}. The selection of bonds can be uniform or random, with the latter referred to as a random-bond AKLT state. Moreover, our proof in Appendix~\ref{app:oldproof} shows that $\otimes_v F_{\alpha_v,v}\otimes_e|{\rm Bell\, state}\rangle_e$ is still an encoded graph state regardless of the labels $\{\alpha_v\}$'s for the POVM outcome and the Bell states used on the edges $e$'s.

In the following, we will discuss the creation of various AKLT states and their deformed version, beginning with one dimension.

\section{Constant-time scheme for the 1D  
 deformed  AKLT states}
 \label{sec:1D}
 

We will first warm up in the 1D case.
 To create the 1D AKLT state, one can prepare a set of singlets along a line and apply the projection $P_v$ to all sites, as defined by the mathematical definition of the AKLT state. This, however, does not work deterministically, as there is an undesired measurement outcome that projects onto the subspace orthogonal to the symmetric one, which is just the singlet. (Note that this effectively removes the two qubits and shrinks the length, as if the two qubits were consumed in performing the entanglement swapping; see,  e.g., a similar scenario in Fig.~\ref{fig:AKLT1D}d.) 
 
 Alternatively, one can create the four-qubit elementary block shown in Fig.~\ref{fig:AKLT1D}b:  $|\Psi_B\rangle\sim P_{23}  |\Psi^-\rangle_{12}\otimes   |\Psi^-\rangle_{34}$. Note that by direction calculation, $|\Psi_B\rangle$ is (up to normalization)
\begin{equation}
|\Psi_B\rangle\sim|0011\rangle+|0101\rangle+|1010\rangle+|1100\rangle-2|0110\rangle-2|1001\rangle.
\end{equation}
 If we have two such blocks and if we can fuse the two dangling qubits (one from each block) by projecting them into a singlet, then we form a bigger AKLT segment, as illustrated in Fig.~\ref{fig:AKLT1D}c by Bell-state measurement (BSM). Similar to the site projection, we cannot guarantee measurement of a singlet, and will with some probability project onto a triplet state. However, it was shown in Ref.~\cite{smith2024constant} that it is possible to correct such undesired outcomes by performing a product of local corrections, exploiting the symmetric properties of the AKLT state. Below, we describe this strategy in full, beginning with the preparation of elementary building blocks.

\subsection{Preparing the elementary block state}
\begin{figure}[h]
    \centering
    \includegraphics[width=\linewidth]{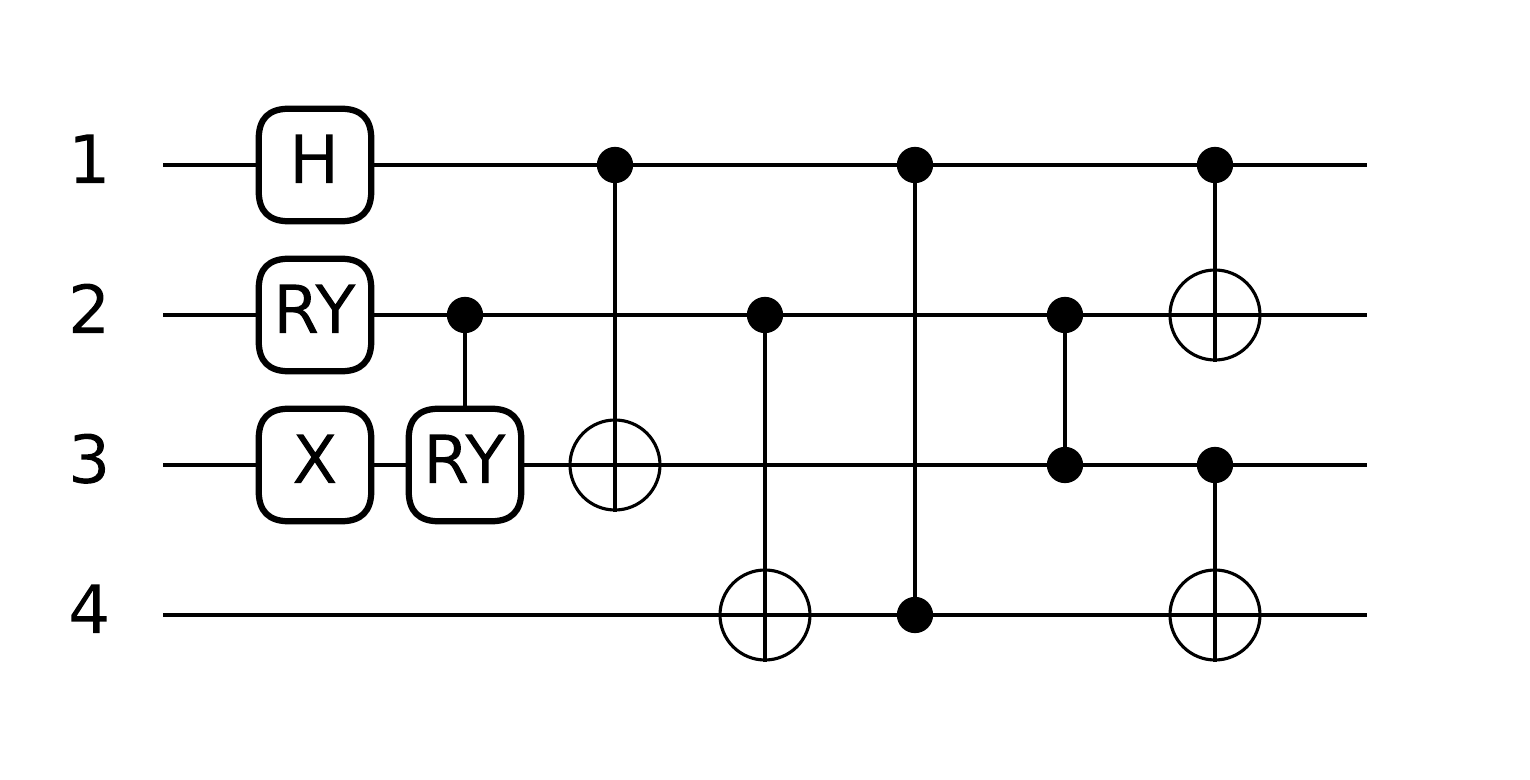}
    \caption{Example circuit that creates $|B(2)\rangle$ using Eq.~(\ref{eq:PsiB1}).}
    \label{fig:B1Circuit}
\end{figure}

The elementary block state  $|\Psi_B\rangle\sim P_{23}  |\psi^-\rangle_{12}\otimes   |\Psi^-\rangle_{34}$  (see Fig.~\ref{fig:AKLT1D}b) can be prepared by the following circuit,
\begin{eqnarray}|\Psi_B\rangle_{1234}&=&{\rm CX}_{34} {\rm CX}_{12} {\rm CZ}_{23} {\rm CZ}_{14} {\rm CX}_{13} {\rm CX}_{24} \nonumber\\
& &H_1 {\rm CR}_{Y,23}(\theta_2) R_{Y,2}(\theta_1) X_3|0000\rangle_{1234},
\label{eq:PsiB1}
\end{eqnarray}
where ${\rm CR}_Y(\theta)=|0\rangle\langle0|\otimes I + |1\rangle\langle1|\otimes R_Y(\theta)$, with  $\theta_2$ satisfying $[\cos(\theta_2/2),\sin(\theta_2/2)]=[2,-1]/\sqrt{5}$,  is a controlled Y rotation,  $R_{Y,2}(\theta_1)$ is a single qubit Y rotation on qubit 2 with $[\cos(\theta_1/2),\sin(\theta_1/2)]=[1,5]/\sqrt{6}$,  $X$ is the Pauli X gate, 
$H$ is the Hadamard gate, and CX and CZ are, respectively, the controlled-X and controlled-Z gates. The circuit for creating this state is shown in Fig.~\ref{fig:B1Circuit}. While this suffices to show that the preparation of the elementary block is both possible and simple, we note that the above is not necessarily the optimal gate sequence. Later, we will explore how to create the general elementary block states for higher dimensional AKLT states, such as those shown in Fig.~\ref{fig:AKLT1D}b.

\subsection{Fusing blocks}

\begin{figure}[h]
    \centering
\includegraphics[width=0.85\linewidth]{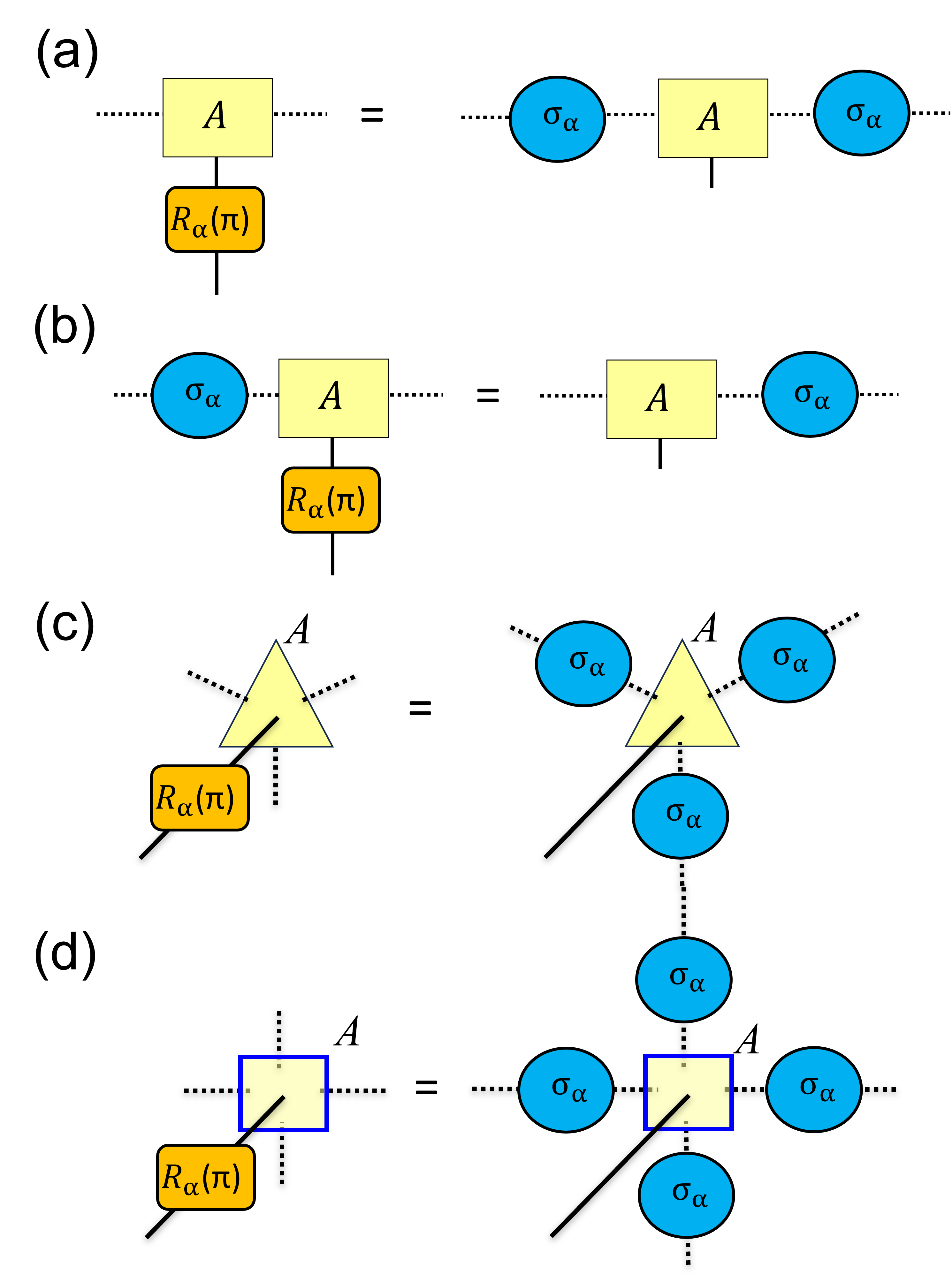}
    \caption{Examples of symmetry in AKLT states. (a) The $Z_2\times Z_2$ symmetry of the 1D AKLT state can be summarized using a local MPS. (b) The diagram in (a) can be interpreted as moving the defect operator $\sigma_\alpha$ from l.h.s. to r.h.s. by acting on the physical degree of freedom by a $\pi$ rotation w.r.t. to $\alpha$-axis. (c) \& (d) represent the $Z_2\times Z_2$ symmetry for hexagonal- and square-lattice AKLT states. The diagrams hold for the specific choice of deformation that preserves the symmetry. }
    \label{fig:AKLTsymmetry.png}
\end{figure}

\noindent {\bf Via Bell-State measurement}. Once we have two such blocks, we can fuse them together using the BSM, as mentioned above. The desired result of the BSM is $|\Psi^-\rangle$, which successfully fuses the two blocks.
Even if the Bell-state measurement results in other non-singlet outcomes then, similar to quantum teleportation, it can be corrected by applying a corresponding Pauli operator $\sigma_\alpha$ on the qubit at either side of the BSM (see Fig.~\ref{fig:AKLT1D}c) and propagating the effect to the other side using the action shown in Fig.~\ref{fig:AKLTsymmetry.png}b. See also Table~\ref{tbl:BellStates} for the origin of the Pauli correction.

 The correction only works if there is no loop, as it is only in this case that the virtual $\sigma_\alpha$ can be propagated to the boundary using local operations. This ability relies on the fact that the AKLT state is an SPT state, in which moving the by-product or defect operator from one end to the other induces a physical action because of the symmetry. 

\smallskip\noindent {\bf Via Hadamard Test}.
To fuse blocks, we used the BSM in the above. But having understood the relation between the HT and BSM, we can also use the HT on the two dangling quibts from two adjacent blocks to implement the fusion.  If the  outcome $|\Psi^-\rangle\langle \Psi^-|$ occurs, we have successfully completed the fusion. If the outcome is $I-|\Psi^-\rangle\langle \Psi^-|$, then we have added a decorative spin-1 site. The upshot is that we can create an AKLT on a graph with a random decoration of spin-1 sites on the edges; see Fig.~\ref{fig:AKLT1D}d.  In addition to the BSM, We shall also use the HT approach in two-dimensional AKLT states below.

As pointed out in Ref.~\cite{smith2022deterministic}, one consequence of using the combination of the HT and BSM is  we can deterministically create a 1D AKLT state with the periodic BC but not with a fixed length. For example, in step 1, we create an open-BC AKLT state of $L$ sites with dangling qubits, which is deterministic. In step 2, to fuse the two ends, we use the HT which projects them to a singlet with probability $\sim 1/4$~\cite{footnote}, thereby producing a periodic-BC AKLT state with $L$ sites. However, with probability $\sim 3/4$, the two qubits are projected to the subspace of triplets, thereby creating one additional site and resulting in a periodic-BC AKLT state with $L+1$ sites.

In the above step 2, we can 
also use the Bell-state measurement instead, and with probability $1/4$, the two end qubits will be projected to $|\Psi^-\rangle$, and thus the two ends will be successfully linked. However, with probability $3/4$, they are projected to triplet states, which corresponds to inserting a Pauli $\sigma$ in the periodic MPS representation of the 1D AKLT state of $L$ sites. This extra Pauli $\sigma$ cannot be removed by unitaries that do not involve all sites, as we prove in Appendix~\ref{app:trapped}. However, we can remove it by measurement at the cost of reducing the number of sites. We measure the physical site next to $\sigma$ in the basis $|x\rangle=(|S_z=+1\rangle+|S_z=-1\rangle)/\sqrt{2}$, $|y\rangle=(|S_z=+1\rangle-|S_z=-1\rangle)/(i\sqrt{2})$,  and $|z\rangle=|S_z=0\rangle$ basis,  
whose effect is to reduce the number of sites by one and insert an outcome-dependent $\sigma_{\alpha=x,y,z}$ in the virtual bond of the corresponding MPS. We can keep measuring the next sites until all the extra Pauli operators acting on the bond cancel (i.e., their product, including the trapped $\sigma$, becomes identity), resulting in a periodic-BC AKLT state, but with a length that cannot be predetermined.

We remark that Ref.~\cite{murta2022preparing} also considers preparing elementary blocks to create AKLT states, but uses the Hadamard test (HT) in order to symmetrize the virtual qubits on physical sites (rather than to  merge blocks) so as to project to the symmetric subspace. (For more than two qubits, projecting to the symmetric subspace requires other more SWAPS between different qubits to implement the projection; see also Ref.~\cite{murta2022preparing}.) In terms of a quantum circuit, one begins with an ancilla qubit initialized in $|0\rangle$, and applies to it a Hadamard gate. Then a conditional SWAP gate is applied on the two target qubits (to be symmetrized) conditioned on the control qubit being $|1\rangle$. Finally, one performs another Hadamard gate on the control qubit and measures it in the $Z$ basis. The circuit is shown below,
\begin{equation}
\Qcircuit @C=0.8em @R=0.0em @!R{  |0\rangle \quad\quad &\gate{H} 
& \ctrl{1}  & \gate{H} & \qw  & \meterB z\\ 
|q_1\rangle \quad \quad & \hphantom{\gate{U}} &\qswap &\qw &\qw \\
|q_2\rangle \quad\quad& \qw &\qswap \qwx&\qw  & \qw &  \\  
\vspace{0em}}
\end{equation}
which is also referred to as a SWAP test.
If the outcome is $|0\rangle$, then the two target qubits are projected into the symmetric subspace, spanned by $|\Phi^\pm\rangle$ and $|\Psi^+\rangle$, equivalently described by the projector $I-|\Psi^-\rangle\langle \Psi^-|$. (Note that, if needed, one can perform a Bell-state measurement to further project on the three basis states.) On the other hand, if the measurement outcome is $|1\rangle$, the two qubits are projected to the singlet $|\Psi^-\rangle$. One thus describes the effect of the HT by a POVM: $\{|\Psi^-\rangle\langle \Psi^-|, I-|\Psi^-\rangle\langle \Psi^-|\}$.  We deterministically create the elementary building block as in Fig.~\ref{fig:AKLT1D}, which contains an effective spin-1 site in the middle. In contrast to Ref.~\cite{murta2022preparing} where the HT is used to probabilistically form the spin-1, this strategy shuffles the problem of creating the AKLT state to how to fuse the blocks, as described earlier. 
 
\subsection{The deformed case}
Our procedure generalizes directly to the deformed AKLT states~\cite{klumper1993matrix}, related to their undeformed counterpart by a local operator $\hat{Q}({{a}})={a}|\underline{+}\rangle\langle\underline{+}|+{a}|\underline{-}\rangle\langle\underline{-}| +|\underline{0}\rangle\langle \underline{0}|$ with $a\in(0,\infty)$: 
\begin{equation}\label{eq:1Ddeformed}|\Psi_{\rm DF}  (a)\rangle=\hat{Q}(a)^{\otimes N}|\Psi_{\rm AKLT}\rangle,
\end{equation}
where $|\Psi_{\rm DF}  (a)\rangle$ is the (un-normalized) deformed AKLT state, reducing to the original AKLT state at $a=1$. Note that $|\underline{\pm}\rangle$ and $|\underline{0}\rangle$ are eigenstates of spin-1 $S_z$ operators. Such a deformation preserves the $Z_2\times Z_2$ symmetry (as well as $U(1)$, which is not important here) in the wavefunction. The deformed  Hamiltonian is $H(a)=\sum_{j} \hat{Q}(a)^{-1}\otimes \hat{Q}(a)^{-1} P^{S=2}_{j,j+1}\,\hat{Q}(a)^{-1}\otimes \hat{Q}(a)^{-1}.$ 
The above steps to create the AKLT state work almost exactly the same way for the deformed case, except that the rotation angles are changed to 
$[\cos(\theta_2/2),\sin(\theta_2/2)]=[2a,-1]/\sqrt{1+4a^2}$,   and $[\cos(\theta_1/2),\sin(\theta_1/2)]=[1,\sqrt{1+4a^2}]/\sqrt{2+4a^2}$.

The fusion and correction procedure works exactly as in the undeformed case, because the deformed AKLT state has the same symmetry, namely $Z_2\times Z_2$ as in Fig.~\ref{fig:AKLTsymmetry.png}a, due to the choice of deformation having the same weights for both $|S_z=+1\rangle$ and $|S_z=-1\rangle$ .

\subsection{String-order parameter}
Regarding the verification, one can measure the local Hamiltonian terms to check the energy. Because of the energy gap of the AKLT model, one can use the local energy and the global gap to bound the fidelity of the measured state with the exact ground state. 
Another useful tool is the string order parameters~\cite{denNijs1989preroughening,girvin1989hidden}, which  probe the SPT order (and the hidden $Z_2\times Z_2$ symmetry breaking)
\begin{equation}
\label{eq:stringorder}
    {\cal O}_{i,i+r}^{\alpha}=-S^\alpha_i e^{i\pi\sum_{j=i+1}^{i+r-1}S_j^\alpha}S^\alpha_{i+r}, \ \alpha=x,y,z,
\end{equation}which can be measured~\cite{smith2022deterministic}. For the above deformed 1D AKLT state, by using the technique of MPS~\cite{perez-garcia2007matrix}, we find that  (see also Appendix~\ref{app:String})
\begin{eqnarray}
    \langle{\cal O}_{i,i+r}^{\alpha}\rangle_\text{AKLT}&=&\frac{4a^4}{(1 + 2a^2)^2}
\end{eqnarray}
is independent of the distance $r$, as is the case at the AKLT point.
 We mention that there are other schemes for verifying AKLT states~\cite{chen2022efficient,zhu2022efficient} that exploit the frustration-free and nonzero-gap properties of AKLT Hamiltonians.

\subsection{1D AKLT quasichain}
\begin{figure}[h]
    \centering
    \includegraphics[width=\linewidth]{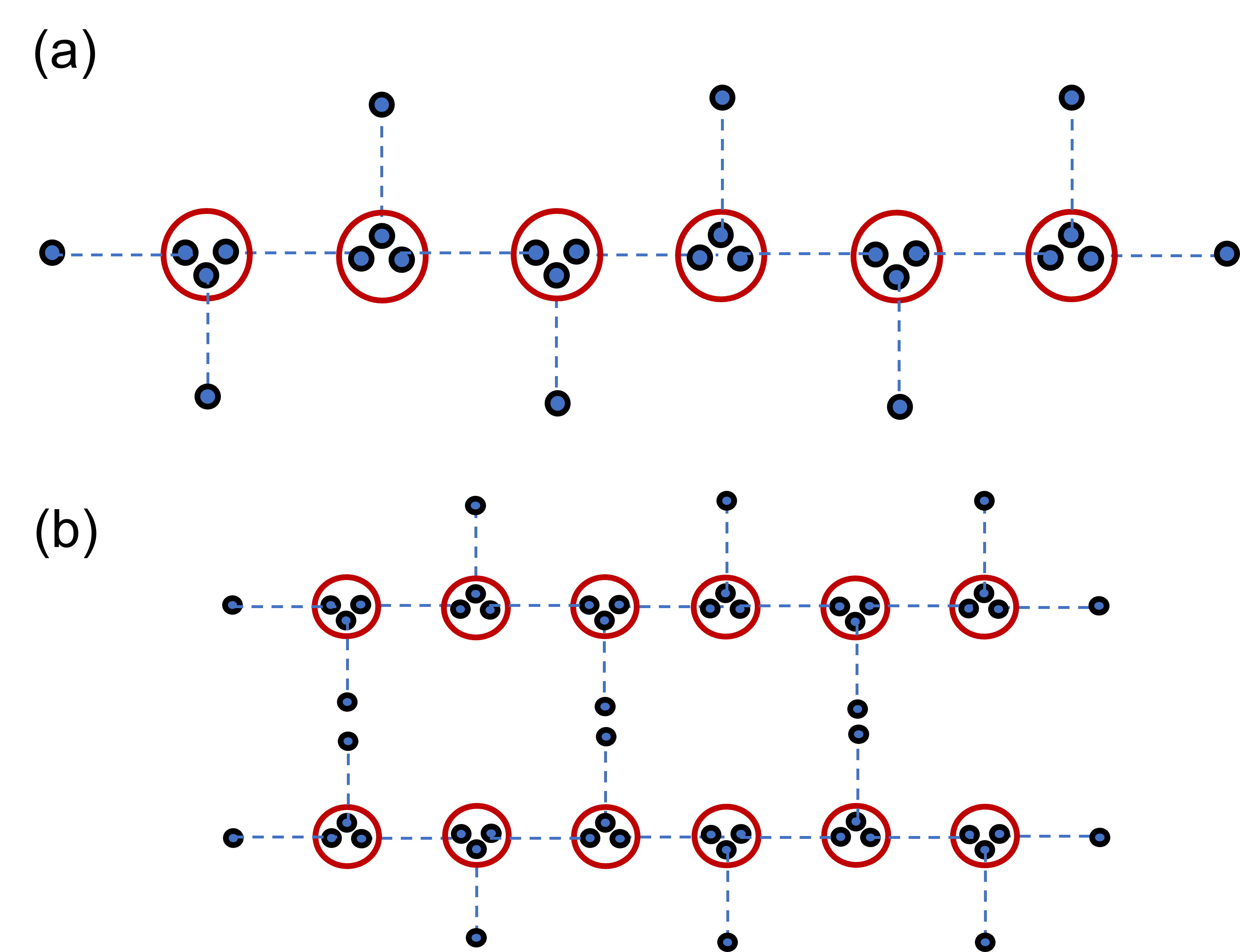}
    \caption{(a) AKLT quasichain that consists of spin-3/2  and spin-1/2 sites. (b) Stack of two such quasichains. This can be used to build a 2D system. There can be several alternatives to merge the neighboring two dangling spin-1/2 sites.  }
    \label{fig:Quasichains}
\end{figure}

As another example in one dimension, we discuss an AKLT quasichain, shown in Fig.~\ref{fig:Quasichains} and first considered in Ref.~\cite{cai2010universal} for MBQC. It consists of spin-3/2 sites in the backbone with additional dangling spin-1/2 sites connected to it. Using the POVM in Eq.~(\ref{eq:POVM}) applied to all spin-3/2 sites, it is easy to see that the post-POVM state will be an encoded 1D cluster state, useful for simulating the operation on one logical qubit~\cite{AKLTquasichain}. Ref.~\cite{cai2010universal} shows that by stacking up such 1D quasichains into a 2D structure (see Fig.~\ref{fig:Quasichains}b for the examples of two stacked chains), local measurement on spin-3/2 sites and joint measurement on neighboring spin-1/2 sites enable universal MBQC.

Our schemes also allow us to build the AKLT quasichain deterministically by (1) creating the spin-3/2 building blocks, as in Fig.~\ref{fig:AKLT1D}b, (2) measuring them via the BSM, (3) correcting them by utilizing the symmetry of the AKLT state to move defects to the boundary, and then correct them.  How to construct the spin-3/2 building blocks and more general ones will be discussed in Sec.~\ref{sec:Blocks}. Their deformed version can also be created similarly.
\section{Creation of (deformed) AKLT states on arbitrary graphs}
\label{sec:arbitrary}

Our approach described above can be generalized to arbitrary graphs and the key ingredient is to create the corresponding elementary building blocks with the correct coordination numbers, as illustrated in Fig.~\ref{fig:AKLT1D}e for the honeycomb ($z_v=3$) and the square ($z_v=4$) lattices. We will systematically construct circuits to prepare these elementary building blocks for arbitrary vertex degrees. One may also consider using larger blocks, as was used in the 1D case~\cite{smith2022deterministic}. The deformed AKLT models in 2D~\cite{NiggemannKlumperZittartz1997,NiggemannKlumperZittartz2000} can possess a richer phase diagram, e.g., for the spin-2 case there is an XY-like phase~\cite{PomataHuangWei18}. We will also work out circuits for creating the deformed building blocks. 

Once we have the building blocks created (done in parallel), we then implement the measurement-based fusion via the Hadamard test or the Bell-state measurement (or combination of them) to merge them. As we have seen in the 1D case, there can be additional decorations of spin-1 sites inserted to edges of the intended lattice on which the AKLT state will reside or the bonds between two spins are not singlets. Then, we need to perform correction, either by measurement or  by exploiting the symmetry of AKLT states. 

However, we are not yet able to create AKLT states deterministically on arbitrary graphs due to loops, which prevent us from pushing the `defects' to the boundary or canceling them within a loop. As we have seen earlier in the 1D case, even though one can create a periodic AKLT state deterministically, the length cannot be predetermined.
As shown in Appendix~\ref{app:trapped}, a defect $\sigma_\alpha$ trapped in the periodic chain cannot be removed by a unitary that does not act on the entire chain. Although it can be removed by measurement, the final length cannot be fixed.  

At this point, it seems that we may not be able to deterministically create a fixed  AKLT state on a graph that has loops using finite-depth adaptive circuits assisted by measurements. 
Nevertheless, we can deterministically create a tree AKLT state, e.g., on a Bethe lattice and its decoration, and there is an interesting order to disorder transition in that system~\cite{AKLT2,Pomata}.

In the following, we first discuss how to create larger building blocks with unitary circuits, and then discuss the resultant states obtained by merging them.  Depending on how we merge them, we can either obtain (1) an AKLT state modified from the desired one with possible edge decorations corresponding to extra spin-1 sites or (2) an AKLT state with random bonds (to be explained below).
\section{Creating general AKLT elementary building blocks and merging them}
\label{sec:Blocks}
As seen in Fig.~\ref{fig:AKLT1D}b, one elementary block consists of $n$ qubits (those inside a circle in the figure) that will be projected to their symmetric subspace and $n$ other ancillary qubits (shown as dangling qubits in the figure). This state will be shown below to be a `doubled Dicke' state $|B(n)\rangle\equiv\sum_k (-1)^k|D(n,k)\rangle |D(n,\bar{k})\rangle/\sqrt{n+1}$, where $\bar{k}=n-k$ and $|D(n,k)\rangle$'s are the so-called Dicke states. Regarding labeling, the first $n$ qubits in $|B(n)\rangle$ represent those inside the circle and the last $n$ qubits represent those dangling qubits.  In this convention, qubits (1234) in $|\Psi_B\rangle$ in Eq.~(\ref{eq:PsiB1}) correspond to qubits (3124) in  $|B(2)\rangle$. 

Now, let us derive the form of the elementary building block state. By definition, it is constructed by applying the total symmetric subspace projector $\hat{P}_S$ over one of the qubits from each of the $n$ Bell pairs in the composite state $|\Psi^-\rangle^{\otimes n}$:
\begin{eqnarray}
    |B(n)\rangle = (\hat P_s)_{1..n}\bigotimes_{i=1}^n |\Psi^-\rangle_{i,i+n},
\end{eqnarray}
where  $|\Psi^-\rangle = \frac{1}{\sqrt 2}(|01\rangle-|10\rangle)$ is the Bell pair, $\hat{P}_s=\sum_{k=0}^n |D(n,k)\rangle\langle D(n,k)|$ projects $n$ qubits to their total symmetric subspace, and we have labeled the projected qubits 1 through $n$ (and their unprojected pair qubits $n+1$ to $2n$).

Carrying out the projection, we get the (unnormalized) state
\begin{eqnarray}
 &&|B(n)\rangle \sim (\hat P_s)_{1..n}\bigotimes_{i=1}^n\frac{1}{\sqrt 2}(|01\rangle_{i,i+n}-|10\rangle_{i,i+n})\\
 &=&\frac{1}{2^{n/2}}Y_{n+1}\dots Y_{2n}(\hat P_s)_{1..n}\bigotimes_{i=1}^n(|01\rangle_{i,i+n}+|10\rangle_{i,i+n}) \nonumber\\
 &=&Y_{n+1}\dots Y_{2n}\sum_k|D(n,k)\rangle_{1,\dots,n}|D(n,k)\rangle_{n+1,\dots,2n}\nonumber\\
 &=&\sum_k (-1)^k|D(n,k)\rangle \otimes |D(n,n-k)\rangle,
\end{eqnarray}
where we have used the identity $(\hat{P}_S)_{1..n} \prod_{i=1}^n (|00\rangle_{i,i+n}+|11\rangle_{i,i+n})=\sum_k|D(n,k)\rangle_{1,\dots,n}|D(n,k)\rangle_{n+1,\dots,2n}$, which can be verified by direct calculation. Applying Pauli $Y$ gates on $|D(n,k)\rangle$ flips its $k$ spin-ups and $\bar{k} = n-k$ spin-downs with a relative phase $(-1)^k$. Restoring the normalization, we arrive at
\begin{eqnarray}
    |B(n)\rangle\equiv\frac{1}{\sqrt{n+1}}\sum_{k=0}^n (-1)^k|D(n,k)\rangle |D(n,\bar{k})\rangle.
\end{eqnarray}

Next, we consider how to prepare such a doubled Dicke state with a short-depth circuit.
To that end, there are recent works~\cite {buhrman2024state,piroli2024approximating} that use mid-circuit measurement and feedforward to create a Dicke state with constant depth. Separately, an earlier work~\cite{bartschi2019determinstic} showed that one can use a $O(n)$-depth unitary-only circuit to create $|D(n,k)\rangle$ from the input $|0^{\otimes n-1}1^{\otimes k}\rangle$ without the need of using ancillas. Here, our goal is the doubled Dicke state, which can be prepared by first targeting
 the superposition state
\begin{equation}
|\psi_0\rangle=\frac{1}{\sqrt{n+1}}\sum_{k=0}^n (-1)^k|0^{\otimes n-k}1^{\otimes k}\rangle\otimes|0^{\otimes k}1^{\otimes n-k}\rangle, 
\end{equation}
followed by a coherent application of two parallel Dicke-state creation circuits. Thus, we shall adopt the method from Ref.~\cite{bartschi2019determinstic}.
The goal of preparing$|\psi_0\rangle$ is then reduced to the preparation of the state
\begin{equation}
\label{eq:phi0}
|\phi_0\rangle\equiv |\tilde{\phi}_0\rangle\otimes|0^{\otimes n}\rangle=\frac{1}{\sqrt{n+1}}\sum_{k=0}^n |0^{\otimes n-k}1^{\otimes k}\rangle\otimes|0^{\otimes n}\rangle, 
\end{equation}
followed by CNOT gates to duplicate the computational-basis information from $i$-th qubit to $(2n-i)$-th qubit and $Z$ gates on qubits $1$ to $n$ (or equivalently on qubits $(n+1)$ to $2n$) in order to obtain the sign $(-1)^k$ and $X$ gates on qubits $(n+1)$ to $2n$ in order to prepare opposite bits from the first group. Namely,
\begin{eqnarray}
\label{eq:psi0phi0}
    |\psi_0\rangle= (-i)^n\prod_{j=1}^n  Y_{j+n} \prod_{i=1}^n {\rm CNOT}_{i,2n-i}|\phi_0\rangle.
\end{eqnarray}
We note that long-distance CNOT gates can be implemented by additional ancillas and dynamical circuits with constant depth~\cite{baumer2024efficient}. If we do not employ ancillas, we can simply use SWAP gates to achieve the long-distance CNOT gates at a cost of longer depth. Given that we are largely focused on building blocks composed of a relatively small number of qubits for AKLT states with small coordination number, we are not worried about the scaling here.

Regarding the preparation of  $|\tilde{\phi}_0\rangle$, which is sparse, we note that, for example, there is a unitary scheme using a circuit with depth $\Theta(\log(nd))$ and $O(nd\log (d))$ ancillas~\cite{zhang2022quantum}, and it applies to our case with $d=n+1$. Here, however, we are also content with a $\Theta(n)$-depth circuit without ancillas, as $n$ will typically not be large. One can verify that the following circuit does achieve the desired creation,
\begin{eqnarray}
   \label{eq:phitilde0} |\tilde{\phi}_0\rangle= \Big[\tprod_{k=n}^{2}{\rm CRy}_{k,k-1}(\theta_{k-1})\Big] \,{\rm Ry}_{n}(\theta_n)|0^{\otimes n}\rangle,
\end{eqnarray}
with ${\rm Ry}$ being the Y rotation, ${\rm CRy}$ the controlled-Ry (with the first index representing the control and the second representing the target), and the angles $\theta_k$ satisfying $\cos(\theta_k/2)=\sqrt{1/(k+1)}$ and $\sin(\theta_k/2)=\sqrt{k/(k+1)}$. We note that $\tprod$ means applying each term in order on the left (time order).

Once we have obtained $|\psi_0\rangle$, we then apply two copies of the unitary circuit in Ref.~\cite{bartschi2019determinstic} on $|\psi_0\rangle$ to create the double Dicke state $|B\rangle$, 
\begin{eqnarray}
  && |B(n)\rangle=  U_D \otimes U_D|\psi_0\rangle
\\
&=& \sum_{k=0}^n (-1)^k(U_D|0^{\otimes n-k}1^{\otimes k}\rangle)\otimes(U_D|0^{\otimes k}1^{\otimes n-k})\rangle  \nonumber \\ 
    &=& \sum_k (-1)^k|D(n,k)\rangle \otimes |D(n,n-k\rangle).
\end{eqnarray}
If $n$ is not large, the creation procedure is still efficient. But further optimization is very likely. 

\begin{figure*}
    \centering
    \includegraphics[width=\linewidth]{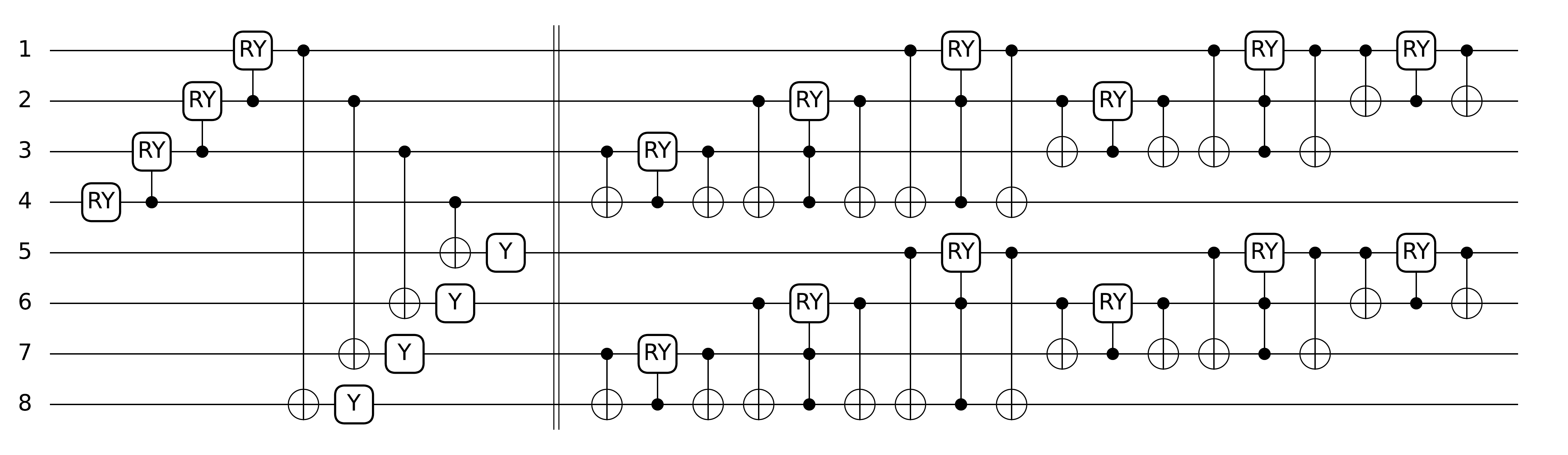}
    \caption{An example circuit that creates $|B(4)\rangle$ by using the Split \& Cyclic Shift method~\cite{bartschi2019determinstic}.  }
    \label{fig:B2Circuit}
\end{figure*}

Next, we summarize the result in Ref.~\cite{bartschi2019determinstic} that constructs  the action of a single copy of $U_D$ that can create $|D(n,k)\rangle$ from $|0^{\otimes n-k}1^{\otimes k}\rangle$ unitarily, for any $k=0,\dots,n$.   
$U_D$ can be constructed by recursively applying the Split \& Cyclic Shift module~\cite{bartschi2019determinstic},
\begin{eqnarray}
   U_D &=&   \tprod_{m=n}^{2} {\rm SCS}_{m,m-1} \\
    {\rm SCS}_{n,k} &=& \tprod_{l=1}^{k} \overline{{\rm SCS}}_{n,l}, \nonumber \\ 
    \overline{{\rm SCS}}_{n,l\geq 1} &=& {\rm CX}_{n-l,n} {\rm CCRy}_{n,n-l+1,n-l}(\theta^{(scs)}_{n,l}) {\rm CX}_{n-l,n}, \nonumber \\
    \overline{{\rm SCS}}_{n,l=1} &=& {\rm CX}_{n-l,n} {\rm CRy}_{n,n-l}(\theta^{(scs)}_{n,l}) {\rm CX}_{n-l,n}, \nonumber \\
    \theta^{(scs)}_{n,l} &=& 2 \arccos \sqrt{\frac{l}{n}}.\nonumber
\end{eqnarray}
The recursive nature of the construction originates from the recurrence of the Dicke states:
\begin{equation}
    |D(n,k)\rangle=\sqrt{\frac{k}{n}}|D(n\!-\!1,k\!-\!1)\rangle|1\rangle +\sqrt{\frac{n\!-\!k}{k}}|D(n\!-\!1,k)\rangle |0\rangle.
\end{equation}
 The total circuit size and depth are $O(n^2)$ and $O(n)$, respectively, by using the method described in~\cite{bartschi2019determinstic}.
An example circuit for creating $|B(n=4)\rangle$ on 8 qubits is demonstrated in Fig.~\ref{fig:B2Circuit}. 
For our purposes, $n$ is not very large, so such a scaling is sufficient. However, we leave the natural question of how to make the entire creation more efficient by other approaches (e.g.,~\cite{buhrman2024state}) for future exploration.

\subsection{Deformed building blocks}

Let us first revisit the basic building block of the deformed $S=1$ AKLT state, which is
\begin{equation}
|0011\rangle+|0101\rangle+|1010\rangle+|1100\rangle-2a|0110\rangle-2a|1001\rangle,
\end{equation}
where $a$ is the deformation parameter.
This can be prepared by Eq.~\eqref{eq:PsiB1} with
\begin{eqnarray}
    \theta_1 &=& 2 \arccos \sqrt{\frac{1}{2 + 4 a^2}},\\
    \theta_2 &=& -2 \arccos \sqrt{\frac{4 a^2}{1 + 4 a^2}}.
\end{eqnarray}

To deal with the case of more general building blocks beyond spin-1, one can consider any local deformation $Q(\{a\})= \sum_{k=0}^n a_k|D(n,k)\rangle\langle D(n,k)|$ such that the deformed AKLT state is $|\psi_{\rm deformed}\rangle=Q(\{a\})^{\otimes N}|\psi_{\rm AKLT}\rangle$. In this case, the building block state is
\begin{equation}
    |\tilde{B}_n(\{a\})\rangle\equiv\frac{1}{\sqrt{\sum_k |a_k|^2}}\sum_k a_k (-1)^k|D(n,k)\rangle |D(n,\bar{k})\rangle.
\end{equation}
This requires one to prepare
\begin{equation}\label{eq:psitlde}
|\tilde{\psi}_0\rangle=\sum_{k=0}^n (-1)^k a_k|0^{\otimes n-k}1^{\otimes k}\rangle\otimes|0^{\otimes k}1^{\otimes n-k}\rangle, 
\end{equation}
where we assume $a_k\ge 0$ in the above, and, in Eq.~(\ref{eq:psitlde}), $\sum_k a_k^2=1$ for simplicity, thus simplifying the denominator ${\sqrt{\sum_k |a_k|^2}}$.
This state can be prepared via the procedure according to Eqs.~(\ref{eq:phi0}),~(\ref{eq:psi0phi0}), and (\ref{eq:phitilde0})
with the rotation angles in Eq.~(\ref{eq:phitilde0}) being replaced by
\begin{eqnarray}
    \theta_{n-l} &=& 2 \arcsin \sqrt {\frac{\sum_{i=l+1}^{n}a_i^2}{\sum_{i=l}^{n}a_i^2}}.
\end{eqnarray}
We remark that as in the 1D case~(\ref{eq:1Ddeformed}), our deformation will be chosen to satisfy $a_k=a_{n-k}$, so as to preserve the $Z_2\times Z_2$ symmetry in the wavefunction (as well as $U(1)$).

\subsection{Deterministic preparation of AKLT states on Bethe lattices}
\label{sec:BetheLattices}

Because they do not contain loops, it is possible to deterministically create AKLT states on undecorated Bethe lattices with any coordination number $z$ (see, e.g., Fig.~\ref{fig:2DAKLT}a) with open boundary conditions, or more precisely, perfectly balanced trees. 
To create the state, one first creates a set of corresponding elementary building blocks with coordination number $z$, as shown in Fig.~\ref{fig:AKLT1D}b, and subsequently merges them via Bell-state measurement, resulting in random Pauli defects along the fused bonds. One then employs the local symmetry of the AKLT state, as shown in Fig.~\ref{fig:AKLTsymmetry.png}cd, by using local $Z_2\times Z_2$ operators to move the defects outward toward the boundary. If the leaves at the boundary are terminated with qubits, then the final correction can be applied to these physical qubits.  Thus, we can create deterministically AKLT states on the Bethe lattices that are terminated with qubits. It is also possible to add spin-1 decorations by employing the HT using the techniques already discussed.

It has been shown that there are interesting order-disorder phenomena in AKLT states~\cite{AKLT2,Pomata} on both the original Bethe lattices and their decorated variants. We can thus employ the techniques discussed here to prepare these states and probe the magnetic order of the states, including possible transitions. In particular, Ref.~\cite{Pomata} shows that with a uniform decoration of $n$ spin-1 sites on the (infinite) Bethe lattice, the AKLT state on such a lattice with coordination number  $z_c=3^{n+1}$ at the branching is a critical state. For $z<z_c$, it is disordered, whereas for $z>z_c$, it is N\'eel ordered.
Realizing these different AKLT states opens up pathways to probe these different magnetic orderings experimentally.

It is straightforward to extend our construction on Bethe lattices to deformed AKLT states that preserve the $Z_2\times Z_2$ symmetry. We note that the order-disorder transition for deformed AKLT states has not yet been studied.

\subsection{Creating AKLT states by fusing the building blocks}
Now we turn to the preparation of AKLT state on general graphs. Once we have the building block states, we can perform the joint measurement on two of the dangling qubits, one from each of the two neighboring blocks, to merge them into a larger AKLT state. As mentioned in the 1D case, there are two types of fusion measurements one can perform: (1) the Hadamard test and (2) the Bell-basis measurement, with the latter being a fine-grained version of the former, as illustrated in Fig.~\ref{fig:AKLT1D}cd.

\smallskip\noindent {\bf Via Hadamard Test}.
Let us consider (1): the effect of the measurement is described by a POVM that projects either to the antisymmetric subspace $\Pi_A=|\Psi^-\rangle\langle \Psi^-|$, or the symmetric subspace $\Pi_S=|\Phi^+\rangle\langle\Phi^+|+|\Phi^-\rangle\langle\Phi^-|+|\Psi^+\rangle\langle \Psi^+|=I-\Pi_A$.

When fusing unconnected portions of the state, there is a probability $P_A=1/4$ that the two dangling qubits will be projected to the antisymmetric state $|\Psi^-\rangle$, meaning that the block states will be linked by a singlet, consistent with the construction of the AKLT states. 

On the other hand, there is a probability $P_S=3/4$ that the two dangling qubits are projected to the symmetric subspace, which is exactly a spin-1 site that appears in the 1d AKLT state. This means that the resultant state in this region contains an AKLT state with two spin-$z/2$ sites linked by a spin-1 site, which serves as a spin-1 decoration of the edge.

We have thus concluded that the fusion procedure by the HT has the same effect as in the 1D case before: we obtain an AKLT state on the original lattice, possibly with a random decoration of a spin-1 site on each edge. 
However, we do not have control over where the decoration will appear.

\begin{figure}
\centering
\includegraphics[width=0.48\textwidth]{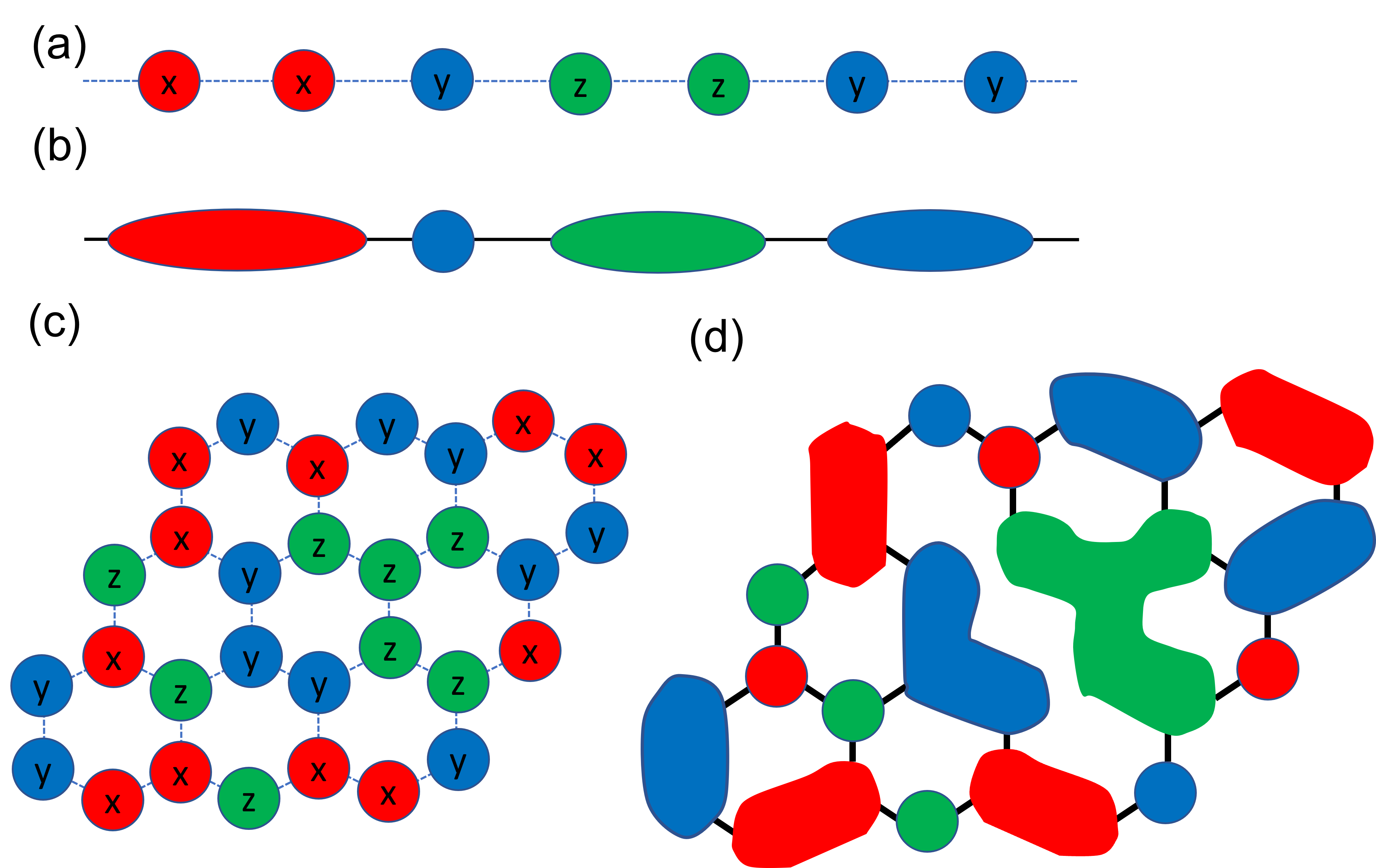}

\caption{Illustration of POVM outcomes and the graph of the post-POVM AKLT graph state. The lower-case x, y, or z  in (a) and (c) indicates the POVM measurement outcome. In (b) and (d), the graphs for the corresponding graph states from performing the POVM on the AKLT states are shown. A vertex can contain a few sites in the original lattice.} 
\label{fig:AKLToutcomesGraph}
\end{figure}

An interesting question is whether we can show that these randomly decorated AKLT states can be used to perform universal measurement-based quantum computation, as the AKLT on the original lattice without any decoration, which seems intuitive. This will be addressed below in Sec.~\ref{ssec:comp_decorated}.

\smallskip\noindent {\bf Via Bell-state measurement}. We could remove the decorations by further performing the Bell-state measurement to reduce the triplet to one of the three Bell states. But this whole process is equivalent to performing the full BSM on the neighboring dangling qubits between two blocks right in the beginning (skipping the HT step). The effect is that the edge linking the two neighboring spin-$z/2$ sites can either be a singlet or any of the triplet Bell states, i.e., there is a random Pauli insertion into the local tensor-network representation of the original AKLT state, similar to the 1D case via the matrix-product-state picture. We call these random-bond AKLT states or AKLT states with random defects. A similar and natural question is whether these random-bond AKLT states are also universal resources for MBQC, as the original AKLT state is. This will be addressed in more specific lattices below.

\section{Quantum computational universality of randomly decorated hexagonal AKLT states and random-bond AKLT states}
\label{sec:Universality}

\begin{figure}
\centering
\includegraphics[width=0.48\textwidth]{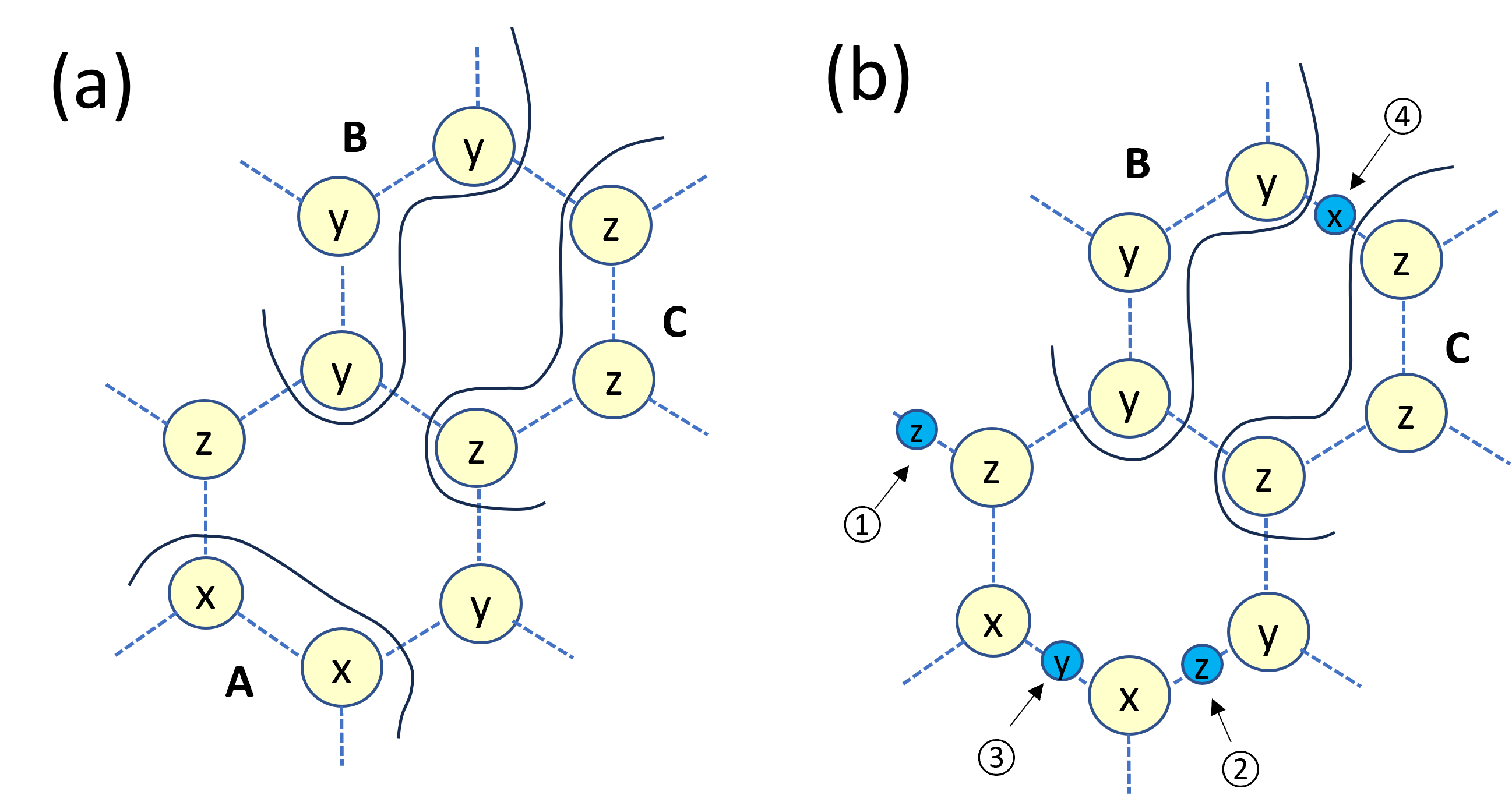}

\caption{Illustration of post-POVM AKLT state. The lower-case x, y, or z indicates the POVM measurement outcome, \textit{not} the correction measurement. (a) An example of measuring AKLT state that has no edge decoration; (b) A corresponding example with some edges decorated and measured. The upper-case letters A, B, and C indicate  (part of the) domains (those connected sites with the same POVM outcome). The numbers inside circles show the cases considered in the main text.} 
\label{fig:PostPOVM_AKLT}
\end{figure}

Let us consider, for illustration of the proof, the hexagonal lattice studied in Ref.~\cite{WeiAffleckRaussendorf11}, as well as other trivalent lattices studied in Ref.~\cite{Wei13}. We first briefly review the argument concluding that the AKLT state on the hexagonal lattice is a universal resource for MBQC. It was shown~\cite{WeiAffleckRaussendorf11} that after a POVM on all spin-3/2 sites, the AKLT state was converted to a random graph state with sufficient connectivity, which can be further reduced to a cluster state, itself a universal resource for MBQC. 

A spin-3/2 state is equivalent to a state in the symmetric subspace of three qubits, the basis states of which are $\{|D(3,k)\rangle\}_{k=0,1,2,3}$. The POVM shown in Eq.~(\ref{eq:POVM}) can be expressed in terms of either picture, and the identity in the spin-3/2 representation $\sum_\alpha F^\dagger_\alpha F_\alpha=I$ is equivalent to the projector onto the three-qubit symmetric subspace.
 The outcomes of the POVM measurement are random (though not completely independent~\cite{WeiAffleckRaussendorf11,wei2012two}); see Fig.~\ref{fig:AKLToutcomesGraph}a,c. 
The meaning of the set of outcomes $\alpha_v$'s on site $v$ is that, for an initial state $|\psi\rangle$, 
the post-measurement state with outcomes $\{\alpha_v\}$ corresponds to $\otimes_v F_{\alpha_v,v}|\psi\rangle$.

For the AKLT state, the post-POVM state was shown to be an encoded graph state, where the encoding means that a vertex of the graph state corresponds to a set of connected sites (which are referred to as a \textit{domain}) that share the same outcome; see, e.g., Fig.~\ref{fig:AKLToutcomesGraph}. Furthermore, the edges of the graph are a subset of the original edges that connect neighboring domains with a special \textit{modulo-2} rule~\cite{WeiAffleckRaussendorf11}: an even number of edges between two domains corresponds to no edge, whereas an odd number of edges between two domains corresponds to a single edge. Intuitively,  this is equivalent to the effect of applying a CZ gate between a pair of qubits an odd or even number of times a CZ gate applies to two qubits. Figures~\ref{fig:AKLToutcomesGraph}b,d illustrate how the graph for the encoded graph state is obtained from the POVM outcome. 
The reason that the AKLT states on the hexagonal and other lattices are universal resources for MBQC is that the post-POVM graph states have graphs that are deep in the percolation phase, and thus they can be converted to a cluster state on a macroscopic-size square lattice~\cite{wei2012two}, the latter being universal for measurement-based quantum computation.
\subsection{Randomly decorated AKLT states}
\label{ssec:comp_decorated}

Let us first consider randomly decorated AKLT states and argue that they are also universal for MBQC. Intuitively, random decorations of spin-1 sites on edges of the original lattice will provide additional logical qubits that can be used for MBQC and should have at least the same quantum computational power as the AKLT state without decoration. Below, we justify such an intuition by the reduction of random graph states. 

Let us begin by considering inserting a spin-1 site on the edge of the two neighboring spin-3/2 sites and performing respective POVMs on them, as described in Eqs.~(\ref{eq:POVMS1}) and~(\ref{eq:POVM}) for spin-1 and spin-3/2, respectively. The POVM outcomes can be characterized as follows. \textcircled{1} The spin-1 site can have the same as either one of the outcome of the two spin-3/2 sites (or both if the two have the same outcomes) or  the spin-1 site have a different outcome from those of spin-3/2 sites: \textcircled{2} when the two spin-3/2 sites have different outcomes;   \textcircled{3} when the two have the same outcome.
These  cases will give different local
graphical structures for the encoded graph state after POVMs. We show these cases in Fig.~\ref{fig:PostPOVM_AKLT}b. Similar to case \textcircled{2}, we also include another example labeled \textcircled{4} there. Case \textcircled{4} exemplifies the scenario that the spin-1 site sits between two extended domains.

The insertion on some edges of the original hexagonal lattice will result in a random graph after the POVM on spin-3/2 sites and the additional spin-1 sites, which is different from the graph associated with one from measuring the AKLT state without decorations (assuming the corresponding spin-3/2 sites on the left and right diagrams in Fig.~\ref{fig:PostPOVM_AKLT} share the same POVM outcomes). Our goal is to perform correction operations to convert the graph on the right diagram to that on the left diagram in Fig.~\ref{fig:PostPOVM_AKLT}.  If this is achievable, it means that the decorated AKLT state can produce the same random graph states as the undecorated AKLT state, implying at least the same quantum computational power.

\begin{figure}
\centering
\includegraphics[width=0.4\textwidth]{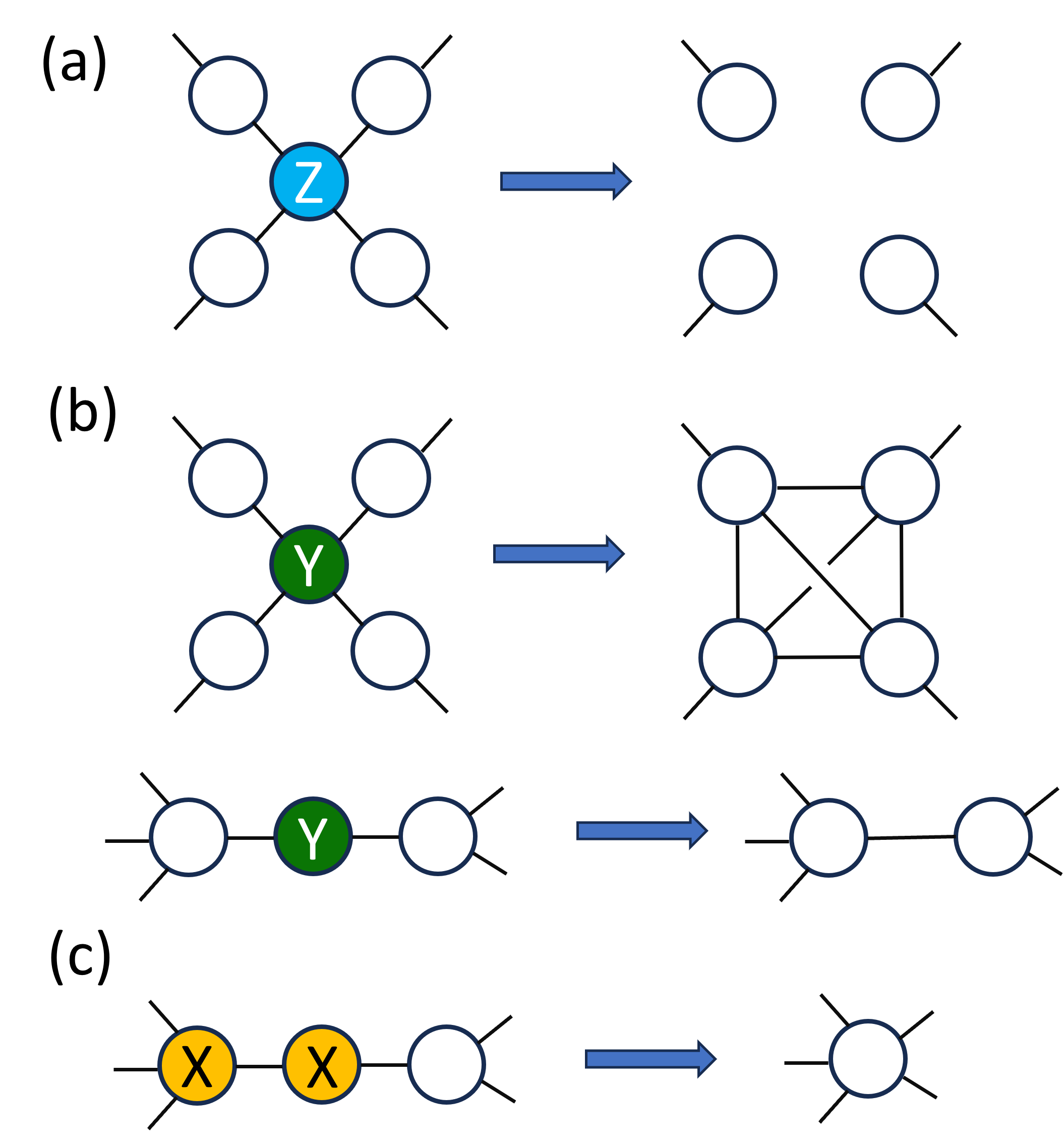}

\caption{Illustration of Pauli Z (a), Y (b), and X (c) measurements on a graph state. The effect is a change of a graph up to local unitaries applied on neighboring qubits. The capital X, Y, and Z symbols indicate what Pauli measurement is performed. } 
\label{fig:PaulMeasuements}
\end{figure}

\begin{table}
  \begin{tabular}{l|r}
  Bell state & Stabilizers \\
  \hline \\
  $|\Phi^+\rangle=-i Y\otimes I|\Psi^-\rangle$ & $X\otimes X$, $-Y\otimes Y$, $Z\otimes Z$ \\
  $|\Phi^-\rangle =-X\otimes I|\Psi^-\rangle$ & $-X\otimes X$, $Y\otimes Y$ , $Z\otimes Z$\\
  $|\Psi^+\rangle=Z\otimes I|\Psi^-\rangle$ & $X\otimes X$, $Y\otimes Y$ , $-Z\otimes Z$\\
  $|\Psi^-\rangle$ & $-X\otimes X$, $-Y\otimes Y$ , $-Z\otimes Z$
  \end{tabular}
 \caption{\label{tbl:BellStates} Bell states and their stabilizer operators. Any of these Bell states can be used as the bond degree of freedom in the construction of the generalized AKLT states, extending the original construction using just the singlet state $|\Psi^-\rangle$. These Bell states share similar stabilizers, up to $\pm$ signs. }
\end{table}

We can recover the original random graph (e.g., the one illustrated in Fig.~\ref{fig:PostPOVM_AKLT}a) from the AKLT by performing additional logical Pauli measurements on the added vertices and possibly neighboring ones.
  For case \textcircled{1}, we do not need to do anything, and the corresponding part of the graph is the same as the undecorated one shown on the left, and it does not matter 
  what the POVM outcome of the spin-3/2 to the left of the spin-1 site (which can be the same or different from that of the right spin-3/2 site). {(Note that the symbol z in Fig.~\ref{fig:PostPOVM_AKLT}b \textcircled{1}, as well as other lower-case letters inside other circles, is the POVM outcome, not the correction measurement.)} 
  For the added vertices in \textcircled{2}, one performs the logical Y  measurement to remove the vertex and thus link their two neighboring vertices; see Fig.~\ref{fig:PaulMeasuements}b for the illustration of the graphical rule corresponding to the (logical) Y-measurement.  For case \textcircled{3}, if the two spin-3/2 sites split by the spin-1 site are still in the same domain due to an extended connection from other parts, this spin-1 site is an isolated vertex (according to the edge modulo-2 rule), so we do not need to do anything. But if the two spin-3/2 sites are now in different domains, then one performs the logical Pauli X
measurement on the spin-1 site vertex and one of its neighbors, which has the effect of merging the two sides of the added vertices;  see Fig.~\ref{fig:PaulMeasuements}c for the illustration of the graphical rule corresponding to the (logical) X-measurement.  For the additional vertex and edge introduced by the decorated site in case \textcircled{4}, we measure the decorated site in the logical Y basis, as in case \textcircled{2}, to link the corresponding sites on both sides and recover the effective edge connection. Because there are two edges between the two domains, the even/odd condition implies that there is no effective edge between them in graph states. (The same correction implemented by the Y measurement applies to the case when there is an odd number of edges between the two domains.)  These exhaust all possible kinds of scenario.

The above procedure can be applied to all random decorated graphs (not just the decorated hexagonal one) to recover the corresponding undecorated graphs. We have also seen that the treatment to recover the graph state in Fig.~\ref{fig:PostPOVM_AKLT}b to that in Fig.~\ref{fig:PostPOVM_AKLT}a is a local measurement that depends only on the information from neighboring sites, except for the information whether a domain is split into two distinct ones. Furthermore, for every configuration on the decorated case, there exists one on the undecorated case that we can recover by local measurement, despite that they emerge from different probability measures.

In the above procedure, we can recover a random graph corresponding to one without decoration. However, before we perform any recovery procedure, the graph from the decorated case generally has more vertices and edges inserted into the former.  In terms of the percolation argument, the probability of having an infinite spanning cluster is not smaller; thus, the graph states have sufficient connectivity to enable at least the same quantum computational power. The reduction we have described in the above procedure rigorously confirms this statement, i.e., if the conversion from one state to another $|\psi\rangle\rightarrow |\phi\rangle$ (up to a product of local unitaries) can be done by measuring qubits locally, then $|\psi\rangle$ has at least the same quantum computational power as $|\phi\rangle$ in the measurement-based approach.
Moreover, for the square lattice and some hybrid ones (involving spin-2 sites), we may encounter non-planar graphs~\cite{WeiPoyaRaussendorf,WeiRaussendorf15}, but the argument of graph reduction above still holds. However, whether random decorations can expand the parameter space of universality is an interesting question for future exploration.

We note that for the deformed and decorated AKLT states, we can also construct their decorated versions. Moreover, there is a certain parameter region for which the deformed AKLT is universal for quantum computation, and, by the above reduction, the decorated case will also be universal.

\begin{table*}
  \begin{tabular}{l|r|r|r}
    \parbox{1.4cm}{POVM outcome\vspace{1mm}} & \multicolumn{1}{c|}{$z$} & \multicolumn{1}{c|}{$x$} & \multicolumn{1}{c}{$y$} \\ \hline
    \parbox{1.6cm}{stabilizer generator} & $\pm Z_iZ_{i+1}$, & $\pm X_iX_{i+1}$ & $\pm Y_iY_{i+1}$ \\
    $\overline{X}$ & $\bigotimes_{j = 1}^{3|{\cal{C}}|} X_j$ or $\bigotimes_{j = 1}^{3|{\cal{C}}|} Y_j$ & $\bigotimes_{j = 1}^{3|{\cal{C}}|} Y_j$ or $\bigotimes_{j = 1}^{3|{\cal{C}}|} Z_j$ & $\bigotimes_{j = 1}^{3|{\cal{C}}|} Z_j$ or $\bigotimes_{j = 1}^{3|{\cal{C}}|} X_j$ \\
    $\overline{Z}$ & $\lambda_i Z_i$ & $\lambda_i X_i$ & $\lambda_i Y_i$ \\
    $\overline{Y}$ & $i \overline{X}\cdot\overline{Z}$ & $i \overline{X}\cdot\overline{Z}$ & $i \overline{X}\cdot\overline{Z}$
  \end{tabular}
 \caption{\label{tbl:coding2} The dependence of stabilizers and encodings for the
 random graphs
  on the local POVM outcome. $|{\cal C}|$ denotes the total number of virtual qubits contained in a
  domain. In the first line, $i = 1\, ..\, 3|{\cal{C}}|-1$,  and in the third line
  $i = 1\,..\, 3|{\cal{C}}|$. The $\pm$ sign in the stabilizer generator is determined from Table~\ref{tbl:BellStates}, according to the underlying Bell state used in the bond. One is free to make the choice of the sign ($\lambda_i$) the logical Z, with the effect being the exchange between the logical 0 and 1 states. It is naturally defined to be the same basis as the POVM outcome. However, there is a freedom in choosing the logical X operator, and once it is decided, the logical Y operator is determined. The choice corresponds to which $b$ we take (by definition). Moreover, for different domains with the same POVM outcome, the choice of the logical X operator can be different. Eventually, we will choose in a way such that the stabilizer operator associated with any domain is $\pm \overline{X}_D \bigotimes_{B \in {\rm Nb}(D)}\overline{Z}_B$.
   }
\end{table*}

\subsection{Random-bond AKLT states}
\label{ssec:comp_randombond}

\begin{figure}
\centering
\includegraphics[width=0.2\textwidth]{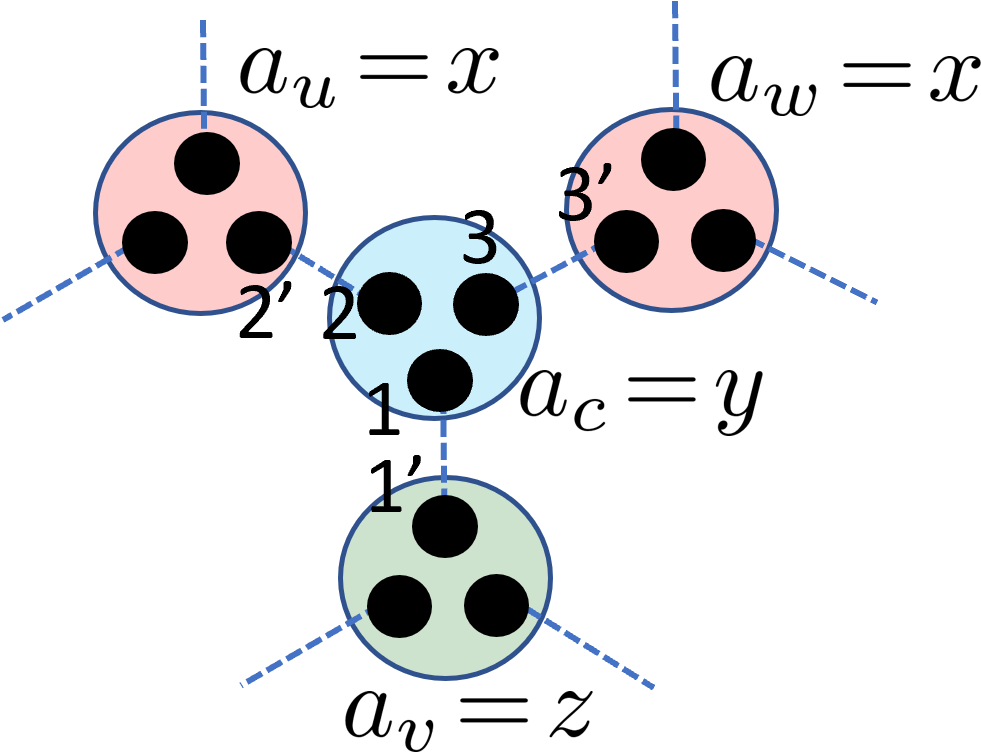}

\caption{Illustration of POVM outcomes (labeled by $a$'s) on four sites $c$, $u$, $v$, and $W$.  The goal is to show that there is a graph-state stabilizer operator centered at site $c$. We will take the logical Z of $u$, $v$, and $w$ to be $Z_u= \sigma_x^{(2')}$, $Z_v= \sigma_z^{(1')}$, and $Z_w= \sigma_x^{(3')}$, respectively, and they commute with  their respective POVM operators $F_x^{(u)}$, $F_z^{(u)}$, and $F_x^{(u)}$. We consider the operator $O_c= [\sigma_z^{(1)}\sigma_x^{(2)}\sigma_x^{(3)}][\sigma_z^{(1')}\sigma_x^{(2')}\sigma_x^{(3')}]$, which is a stabilizer operator (up to a sign) of the three Bell bonds: (1,1'), (2,2'), and (3,3'), irrespective of their type; see Table~\ref{tbl:BellStates}. } 
\label{fig:GraphStabilizer}
\end{figure}

We now  prove the universality of random bond AKLT states.
This case may seem slightly more complicated than the randomly decorated AKLT states considered in the previous subsection, as here the Bell-state measurement gives four different outcomes. The case of the singlet is the desired outcome. The other Bell states correspond to three particular measurement outcomes in the triplet or the equivalent spin-1 subspace on the effective decorated site. (That is we treat the edge with a triplet bond as if there was an effective spin-1 decoration followed by the BSM.) In terms of the graph states, these correspond to some logical Pauli X or Y basis measurement on the effective decorated sites with specific outcomes. But as to which basis X or Y, we do not have any control (see Ref.~\cite{WeiRaussendorf15} for a similar scenario in the spin-2 case).  For the Y measurement, it is fine, as the resultant graph is still planar; see Fig.~\ref{fig:PaulMeasuements}b. But for X, we have to perform additional X measurements to merge the two sides; see Fig.~\ref{fig:PaulMeasuements}c. In the end, we still obtain a random graph state but its graph percolation properties are modified from the corresponding one without random bonds, unlike the case of random decorations. One could still perform Monte Carlo simulations to verify percolation. However, we can alternatively bypass the above issues of determining the logical measurements and performing simulations, and we directly ask whether the post-POVM state on the random bond AKLT is already a graph state in some basis. The answer is indeed positive.

\smallskip
\noindent \textit{Encoded graph states}. It turns out that we can directly translate the proof for the regular AKLT state with singlet bonds~\cite{wei2012two} to the random bond states and show that these random bond AKLT states after the POVM will produce random graph states. This is done in Appendix~\ref{app:oldproof}.
Instead of presenting the full detailed proof here, we will illustrate with an example shown in Fig.~\ref{fig:GraphStabilizer}. Before we move on, we note that we will alternate the symbol for Pauli matrice between $\{\sigma_x,\sigma_y,\sigma_z\}$ and $\{X,Y,Z\}$, with the latter sometimes used for logical Pauli operators. According to the POVM outcomes,
we will take the logical Z of $u$, $v$, and $w$ to be $Z_u= \sigma_x^{(2')}$, $Z_v= \sigma_z^{(1')}$, and $Z_w= \sigma_x^{(3')}$, respectively, as they commute with their respective POVM operators $F_x^{(u)}$, $F_z^{(u)}$, and $F_x^{(u)}$. Next, we consider the operator $O_c= [\sigma_z^{(1)}\sigma_x^{(2)}\sigma_x^{(3)}][\sigma_z^{(1')}\sigma_x^{(2')}\sigma_x^{(3')}]=[\sigma_z^{(1)}\sigma_x^{(2)}\sigma_x^{(3)}]Z_u Z_v Z_w$, which is easily seen to be a stabilizer operator (up to a sign) of the three Bell bonds: (1,1'), (2,2'), and (3,3'), irrespectively of their type; see Table~\ref{tbl:BellStates}. 
Since $a_c=y$, we need to check whether $O_c$ commutes with $F_y$. Since $F_Z^{(c)}\sim(I+ \sigma_y^{(1)}\sigma_y^{(2)})(I+ \sigma_y^{(2)}\sigma_y^{(3)})$, it suffices to show that $[\sigma_z^{(1)}\sigma_x^{(2)}\sigma_x^{(3)}]$ commutes with $\sigma_y^{(1)}\sigma_y^{(2)}$ and $\sigma_y^{(2)}\sigma_y^{(3)}$, which is indeed the case. Thus, $O_c$ is a stabilizer of the post-POVM state.

Next, we will show that the product $[\sigma_z^{(1)}\sigma_x^{(2)}\sigma_x^{(3)}]$ is either a logical X or Y for site $c$, up to a sign.   According to Table~\ref{tbl:coding2},  we can choose
either $X_c=\sigma_z^{(1)}\sigma_z^{(2)}\sigma_z^{(3)}$ or $X_c=\sigma_x^{(1)}\sigma_x^{(2)}\sigma_x^{(3)}$ as the logical X (up to a sign). If we choose the former, then $[\sigma_z^{(1)}\sigma_x^{(2)}\sigma_x^{(3)}]=-X_c \,\sigma_y^{(2)}\sigma_y^{(3)}$. Moreover, the logical Z operator on site $c$ is $Z_c=\sigma_y^{(1)}=\sigma_y^{(2)}=\sigma_y^{(3)}$. Thus,  
$[\sigma_z^{(1)}\sigma_x^{(2)}\sigma_x^{(3)}]=-X_c Z_c^2=-X_c$. On the other hand, if we choose $X_c =\sigma_x^{(1)}\sigma_x^{(2)}\sigma_x^{(3)}$,  then $[\sigma_z^{(1)}\sigma_x^{(2)}\sigma_x^{(3)}]=-iX_c \sigma_y^{(1)}=-i X_c Z_c=Y_c$. Given that we have the freedom to choose which encoding is the logical X, we can always make $O_c=\pm X_c Z_u Z_v Z_w$ (up to a sign), the graph-state stabilizer operator. 

\smallskip
\noindent \textit{Quantum computational universality}.
As shown previously in Refs.~\cite{WeiAffleckRaussendorf11,wei2012two,Wei13,WeiPoyaRaussendorf,WeiRaussendorf15}, the random graphs resulting from the POVM on the various uniform singlet-bond AKLT states reside in the percolation phase. Now we argue that the potential frustration that arises due to the random bond will not move the ensemble of the random graphs across the phase transition to the unpercolated phase. 

The frustration arises when one considers the configuration of POVM outcomes on a loop. Those with identical outcomes (either x, y or z) along a closed path may not be possible due to frustration. Take, for example, a hexagon and take a specific random bond choice with one being $|\Psi^-\rangle$ and the other five being $|\Phi^+\rangle$; see Table~\ref{tbl:BellStates} for their stabilizers. This means that the POVM outcome on the six sites along the hexagon being all $x$ or $z$ is frustrated (i.e., an odd number of stabilizers with minus signs) and thus forbidden. But all y outcomes are allowed. 
One can loosely assume x, y, or z appears with 1/3 without worrying about the correlated outcomes (as the correlated effect is a higher order correction), and conclude that frustration, in this case, appears approximately with a probability of 
$p_{\rm f}=2\times (1/3)^6\approx 2.7\times 10^{-3}$. If we treat the forbidden configuration as a domain to be deleted from the random graph without frustration with probability $p_{\rm f}$, then such a small probability will not take a graph inside the percolated phase across the phase transition. (Moreover,  not all domains are formed by loops of hexagons, so there are other domains that are never deleted, so we have overestimated the average deletion probability in treating the frustration.)  In general, such a frustration is suppressed by a factor $(1/3)^6$ due to the loop size in the hexagonal lattice.

  We can also argue that, for other lattices (such as the square-octagon and cross lattices), the probability of frustration is small, e.g., of order $(1/3)^4$. This probability value is indeed substantially below the domain-deletion threshold (e.g., around 0.3). We note that a similar argument was used to claim the existence of a computational backbone~\cite{Miyake11}, where the difference of the occupation probability $p=2/3$ and the threshold $p_c=1-2\sin(\pi/18)\approx 0.652$ in that case is about 0.014. The difference in the two probabilities we have here is greater than 0.30, as the threshold deletion probability was found in Ref.~\cite{Wei13}. Hence, our argument that the random graphs are in the percolated phase is robust. Moreover, despite the fact that the correlated effect can change the probability $p_f$ by a small fraction, it cannot overcome such a large gap between the two probabilities, so the random graphs remain deep in the percolated phase.

From the above arguments and estimation, we conclude that the {\it random-bond} AKLT states on the lattices where the \textit{original} AKLT states are universal for quantum computation are themselves also universal.
Now that we have concluded the quantum universality for the random-bond case, it can be straightforwardly applied to establish the same universality for the random decorated case, as in the latter case, we just need to perform the fine-grained Bell-state measurement on the decorative spin-1  site (two effective qubits), and the result becomes the random-bond case.

We further comment that it is easy to extend the argument to the deformed random-bond AKLT case (that preserves the $Z_2\times Z_2$ symmetry) and conclude that it should also exhibit quantum-computational universality, i.e., the random-bond case should also have universal computational power as long as the corresponding non-random-bond but deformed case does in the specific parameter range that results in random graphs deep in the percolated phase.

\section{Conclusion}
\label{sec:Conclude}
We consider the problem of creating AKLT states via measurement-based fusion, using local operations (unitary gates, measurements, and feedforward) acting on a few sites. The scheme is deterministic when there is no loop, such as in the 1D open chain or the Bethe lattices. When there is a loop, the state we can create by our scheme is a randomly decorated version (or alternatively a random-bond version) of the target state, where every edge of the target state is decorated with a spin-1 site with probability 3/4.  However, such random decoration does not reduce the power of universal computation from the target state. 
We also consider the case in which the AKLT state is built not only from singlet bonds but also from triplet entangled pairs, which we refer to as random-bond AKLT states. These are derived either by using BSM to fuse elementary building blocks or by measuring the randomly decorated sites. We show that after the POVM on all sites, the resultant state is an encoded graph state, just as in the case of AKLT states with uniform singlets as bonds.

We also argue that both the decorated AKLT states and the random-bond AKLT states possess quantum computational power similar to the undecorated or the uniform singlet-bond AKLT states on the undecorated lattices.
For the singlet-bond AKLT state on the star lattice, it was previously shown that the POVM leads to random graphs that are not in the percolated phase~\cite{Wei13}, thus not suitable for universal MBQC. But with partial decoration, it became universal, as shown in Ref.~\cite{WeiPoyaRaussendorf}. We expect that both the random decoration and the random bonds may help to make the resultant AKLT states on the star lattice universal.   Adapting the argument in Ref.~\cite{Wei13} on the singlet-bond AKLT state on the star lattice (that uses a mapping to the kagom\'e lattice) to our random-bond case, we find that the frustrations leads to an average probability of deleting an edge $p^{\rm[bond]}_{\rm del}\approx 0.4712$ to remove an edge in a triangle on the kagom\'e lattice, which has a bond percolation  $p^{\rm[bond]}_{\rm th}\approx 0.5244$. We see that the probability that a bond is occupied is $p^{\rm[bond]}_{\rm occupied}=1-p^{\rm[bond]}_{\rm del}\approx 0.5288 >p^{\rm[bond]}_{\rm th}$. Thus, it is possible that the random-bond AKLT state on the star lattice after POVM is universal, which, if true, will imply the quantum computational universal for the randomly decorated AKLT state on the star lattice as well. However, the difference of the two probabilities ($p^{\rm[bond]}_{\rm occupied}-p^{\rm[bond]}_{\rm th}$) is about 0.0044, much smaller than those we have for other lattices above. To confirm this requires Monte Carlo simulations to test the percolation property, which is beyond the scope of this paper and is left for future study. 

From the perspective of universal quantum computational power provided by these decorated or random-bond AKLT states, one sees that our constant-time fusion schemes are able to prepare in constant time quantum resource states that sampling generic local measurements on them is already BQP-hard. It would be interesting to utilize random AKLT states to develop protocols that are classically intractable but verifiable, similar to those instantaneous quantum polynomial circuits.

The schemes we propose here, although seemingly discussed in the context of a single quantum processor, can apply to a distributed setting.
Each $2n$-qubit elementary building block retains the center $n$ qubits (that will form an effective spin-$n/2$ site) at a node but sends each of the dangling qubits to its neighboring $n$ nodes (or mid-stations) for the Hadamard test or Bell-state measurement to merge these blocks. Thus, we can obtain a distributed, randomly decorated AKLT state or a distributed random-bond AKLT state. These states can then be used as a distributed resource for quantum computation.

Although we have focused primarily on one- and two-dimensional systems, our approach applies to AKLT states in three or higher dimensions. For example, the 3D AKLT Hamiltonian on the decorated diamond lattice (with a single decorative spin-1 site on each edge) was shown to be gapped~\cite{guo2021nonzero} and thus the AKLT state is a stable disordered state, in contrast to the one on the 3D cubic lattice. Our scheme can be used to create AKLT states on the diamond lattice with decorations on each edge, or a random-bond AKLT state on the diamond lattice (with or without decorations). 

Finally, we conjecture that it is not possible to create 2D or higher-dimensional AKLT states with an extensive number of loops in constant time with finite probability (i.e., that does not depend on the system size) using only short-distance gates and mid-circuit measurements, at least with fusion-based techniques.  In contrast, it would be interesting to explore what classes of 2D tensor-network states can be efficiently prepared~\cite{sahay2024finitedepthpreparationtensornetwork}.  Our approach of creating building block states and then fusing them via the HT or BSM can be extended to other kinds of 2D or higher-dimensional tensor-network states. However, the resulting states may not be correctable to the desired target tensor-network state during merging; instead, they may constitute an ensemble of random states, as in the case of our randomly decorated and random-bond AKLT states. It would be interesting to explore whether the majority of states in such an ensemble can serve as some useful information processing purposes. 

\acknowledgments
This work was primarily supported by the
U.S. National Science Foundation (NSF) under Award No. PHY
2310614. K.S. and T.-C. W. were also supported by the U.S. Department of Energy (DOE), Office of Science, National Quantum Information Science Research Centers, Co-design Center for Quantum Advantage under Contract No. DE-SC0012704. T.-C. W. also acknowledges the support from the Center for Distributed Quantum Processing at Stony Brook University. The part supported by DOE focuses on the algorithmic creation of the elementary block states and that by NSF focuses on the computational universality of random AKLT states.

\appendix

\section{String order parameter for the deformed AKLT state} \label{app:String}

Here, we compute the string order parameters for the 1D deformed AKLT state.
We begin with the MPS for the 1D AKLT state, 
\begin{eqnarray}
    A_0=\sigma_z/\sqrt{2},\;A_{+1}=-\sigma^-,\;A_{-1}=\sigma^+,
\end{eqnarray}
which can be conveniently written in a matrix that contains states in its elements:
\begin{eqnarray}
A = \sqrt{\frac{2}{3}}\begin{bmatrix} |0\rangle/\sqrt{2} & |-1\rangle \\ -|+1\rangle & -|0\rangle/\sqrt{2} \end{bmatrix}.\end{eqnarray}
The deformed AKLT state in Eq.~(\ref{eq:1Ddeformed}) is straightforwardly written as
\begin{eqnarray}
A = \sqrt{\frac{2}{1+2a^2}}\begin{bmatrix} |0\rangle/\sqrt{2} & a |-1\rangle \\ -a|+1\rangle & -|0\rangle/\sqrt{2} \end{bmatrix}.\end{eqnarray}
For convenience, we write down the expression for the string order parameters,
\begin{equation*}
    {\cal O}_{i,i+r}^{\alpha}=-S^\alpha_i e^{i\pi\sum_{j=i+1}^{i+r-1}S_j^\alpha}S^\alpha_{i+r}, \ \alpha=x,y,z,
\end{equation*}
where
\begin{eqnarray}
\!\!\!\!\!\!S_x &=& \frac{1}{\sqrt{2}}
\begin{bmatrix}
0 & 1 & 0 \\
1 & 0 & 1 \\
0 & 1 & 0
\end{bmatrix}, 
\quad e^{\,i\pi S_x}=
\begin{bmatrix}
0&0&-1\\
0&-1&0\\
-1&0&0
\end{bmatrix},\\
\!\!\!\!\!\!S_y &=& \frac{1}{\sqrt{2}}
\begin{bmatrix}
0 & -i & 0 \\
i & 0 & -i \\
0 & i & 0
\end{bmatrix}, 
\quad e^{\,i\pi S_y}=
\begin{bmatrix}
0&0&1\\
0&-1&0\\
1&0&0
\end{bmatrix},\\
\!\!\!\!\!\!S_z &=& 
\begin{bmatrix}
1 & 0 & 0 \\
0 & 0 & 0 \\
0 & 0 & -1
\end{bmatrix}, \quad 
    e^{\,i\pi S_z}=
\begin{bmatrix}
-1&0&0\\
0&1&0\\
0&0&-1
\end{bmatrix}.
\end{eqnarray}

The corresponding transfer matrices are:
\begin{eqnarray}
    M=\sum_i A_i\otimes A_i^* = \frac{2}{1 + 2a^2}
\begin{pmatrix}
\frac{1}{2} & 0 & 0 & a^2 \\
0 & -\tfrac{1}{2} & 0 & 0 \\
0 & 0 & -\tfrac{1}{2} & 0 \\
a^2 & 0 & 0 & \tfrac{1}{2}
\end{pmatrix}, 
\end{eqnarray}
\begin{eqnarray}
    M_{z}&=&\sum_{i,j} (S_z)_{ij} A_i\otimes A_j^* \nonumber\\
    &=& \frac{2}{1 + 2a^2}
\begin{pmatrix}
0 & 0 & 0 & -a^2 \\
0 & 0 & 0 & 0 \\
0 & 0 & 0 & 0 \\
a^2 & 0 & 0 & 0
\end{pmatrix},
\end{eqnarray}
\begin{eqnarray}
    M_{U_z}&=&\sum_{i,j} (e^{\,i\pi S_z})_{ij} A_i\otimes A_j^* \nonumber\\
   & =& \frac{2}{1 + 2a^2}
\begin{pmatrix}
\frac{1}{2} & 0 & 0 & -a^2 \\
0 & -\tfrac{1}{2} & 0 & 0 \\
0 & 0 & -\tfrac{1}{2} & 0 \\
- a^2 & 0 & 0 & \tfrac{1}{2}
\end{pmatrix},
\end{eqnarray}
\begin{eqnarray}
    M_x&=&\sum_{i,j} (S_x)_{ij} A_i\otimes A_j^* \nonumber\\
    &=& \frac{2}{1 + 2a^2}
\begin{pmatrix}
0 & \tfrac{1}{2}a & \tfrac{1}{2}a & 0 \\
-\tfrac{1}{2}a & 0 & 0 & -\tfrac{1}{2}a \\
-\tfrac{1}{2}a & 0 & 0 & -\tfrac{1}{2}a \\
0 & \tfrac{1}{2}a & \tfrac{1}{2}a & 0
\end{pmatrix},
\end{eqnarray}
\begin{eqnarray}
    M_{U_x}&=&\sum_{i,j} (e^{\,i\pi S_x})_{ij} A_i\otimes A_j^* \nonumber \\
    &=& \frac{2}{1 + 2a^2}
\begin{pmatrix}
-\tfrac{1}{2} & 0 & 0 & 0 \\
0 & \tfrac{1}{2} & a^2 & 0 \\
0 & a^2 & \tfrac{1}{2} & 0 \\
0 & 0 & 0 & -\tfrac{1}{2}
\end{pmatrix},
\end{eqnarray}
\begin{eqnarray}
    M_y&=&\sum_{i,j} (S_y)_{ij} A_i\otimes A_j^* \nonumber\\
    &=& \frac{2}{1 + 2a^2}
\begin{pmatrix}
0 & -\tfrac{i}{2}a & \tfrac{i}{2}a & 0 \\
-\tfrac{i}{2}a & 0 & 0 & -\tfrac{i}{2}a \\
\tfrac{i}{2}a & 0 & 0 & \tfrac{i}{2}a \\
0 & -\tfrac{i}{2}a & \tfrac{i}{2}a & 0
\end{pmatrix},
\end{eqnarray}
\begin{eqnarray}
    M_{U_y}&=&\sum_{i,j} (e^{\,i\pi S_y})_{ij} A_i\otimes A_j^*\nonumber\\
    &=& \frac{2}{1 + 2a^2}
\begin{pmatrix}
-\tfrac{1}{2} & 0 & 0 & 0 \\
0 & \tfrac{1}{2} & -a^2 & 0 \\
0 & -a^2 & \tfrac{1}{2} & 0 \\
0 & 0 & 0 & -\tfrac{1}{2}
\end{pmatrix}.
\end{eqnarray}
The infinite-size environment tensor for the periodic boundary condition is
\begin{eqnarray}
    M_\text{env} = \lim_{r\rightarrow\infty}\frac{M^r}{\mathrm{Tr} M^r} = \begin{pmatrix}
\frac{1}{2} & 0 & 0 & \tfrac{1}{2} \\
0 & 0 & 0 & 0 \\
0 & 0 & 0 & 0 \\
\tfrac{1}{2} & 0 & 0 & \tfrac{1}{2}
\end{pmatrix}.
\end{eqnarray}
The expectation values of string order parameters of length $r$ along the respectively axes $\alpha=x,y,z$ of the deformed 1D AKLT spin-chain with infinite size are
\begin{eqnarray}
    \langle{\cal O}_{i,i+r}^{z}\rangle_\text{AKLT}&=&\frac{-\mathrm{Tr}  M_\text{env} M_{z} M_{U_z}^{r-2} M_{z} }{\mathrm{Tr} M_\text{env} M^r } \\
    &=& \frac{4a^4}{(1 + 2a^2)^2},
\end{eqnarray}
\begin{eqnarray}
   \langle {\cal O}_{i,i+r}^{x}\rangle_\text{AKLT}&=&\frac{-\mathrm{Tr}  M_\text{env} M_{x} M_{U_x}^{r-2} M_{x} }{\mathrm{Tr} M_\text{env} M^r } \\
    &=& \frac{4a^4}{(1 + 2a^2)^2},
\end{eqnarray}
\begin{eqnarray}
    \langle{\cal O}_{i,i+r}^{y}\rangle_\text{AKLT}&=&\frac{-\mathrm{Tr}  M_\text{env} M_{y} M_{U_y}^{r-2} M_{y} }{\mathrm{Tr} M_\text{env} M^r } \\
    &=& \frac{4a^4}{(1 + 2a^2)^2}.
\end{eqnarray}
Note that although the deformation breaks the $SO(3)$ rotational symmetry to $U(1)$, the string order parameter expectation values are equal along the three axes.

\section{Proof of Nonexistence of Quasilocal Unitary that can remove a random bond in an AKLT loop}
\label{app:trapped}
Here, we show that there are no quasi-local unitary operators that can remove a defect, i.e., correct the random bond in a spin-1 AKLT loop.
The AKLT state on a loop is
\begin{eqnarray}
    \psi^{(N,I)} = \sum_{s=0,\pm1} \text{Tr}(A_{s_1}A_{s_1}\dots A_{s_N})|s_1s_2\dots s_N\rangle,
\end{eqnarray}
while the one with a random bond or defect is
\begin{eqnarray}
    \psi^{(N,\sigma)} = \sum_{s=0,\pm1} \text{Tr}(\sigma A_{s_1}A_{s_1}\dots A_{s_N})|s_1s_2\dots s_N\rangle,
\end{eqnarray}
where $\sigma = \sigma_x, \sigma_y$, or $ \sigma_z$ is a defect inserted in the bond between, e.g., site $N$ and site 1.
Note that because of the underlying $Z_2 \times Z_2$ SPT order of the 1D AKLT state, all the defects in the random bonds can be moved to the first bond and merged into a single $\sigma$, up to a local unitary transformation.

\smallskip
\noindent Sketch of the proof:
A quasi-local operator can at most act on $N-1$ sites (otherwise it is a global unitary transformation). If we take the density matrix of the remaining site, then it does not change under the quasi-local unitary, 
\begin{eqnarray}
   \rho_1=&&\text{Tr}_{s_2\dots s_N}\big(U_{s_2\dots s_N}|\psi^{(N,\sigma)}\rangle\langle\psi^{(N,\sigma)}|U_{s_2\dots s_N}^\dagger \big)\nonumber \\
 &&  =
    \text{Tr}_{s_2\dots s_N}|\psi^{(N,\sigma)}\rangle\langle\psi^{(N,\sigma)}|. 
\end{eqnarray}
So, if there exists a quasi-local unitary that acts on at most $N-1$ sites (from 2 to $N$) that transforms $|\psi^{(N,\sigma)}\rangle$ into $|\psi^{(N,I)}\rangle$, then the density matrix of the two states on the site $s_1$ must be the same.
\begin{eqnarray}
\!\!\!\!    \text{Tr}_{s_2\dots s_N}|\psi^{(N,\sigma)}\rangle\langle\psi^{(N,\sigma)}| = \text{Tr}_{s_2\dots s_N}|\psi^{(N,I)}\rangle\langle\psi^{(N,I)}|.
\end{eqnarray}

In the next subsection we show that the density matrix of the X, Y, and Z decorated AKLT state on a bond (e.g., between sites 1 \& 2) is different from the density matrix of the original AKLT state. 
Thus, we conclude that we cannot remove a random bond on a spin-1 AKLT loop using quasi-local operators.

\subsection{Density matrix of random bond AKLT state on a single site}

In this subsection, we shall compute the density matrix of the random-bond AKLT state on a loop of $N$ sites, for which we have moved all the random Pauli defects to the bond between site $N$ and $1$ 

The transfer matrix of the AKLT MPS for sites $2-N$ is:
\begin{eqnarray}
    M^{(N)} &=& \sum_{s=0,\pm1} \prod_{i=1}^{N-1} A_{s_i} \otimes A^*_{s_i} \\
    &=&\left(\frac{2}{3}
\begin{pmatrix}
\frac{1}{2} & 0 & 0 & 1 \\
0 & -\tfrac{1}{2} & 0 & 0 \\
0 & 0 & -\tfrac{1}{2} & 0 \\
1 & 0 & 0 & \tfrac{1}{2}
\end{pmatrix}\right)^{N-1}\\
&=&\begin{pmatrix}
\frac{1+q}{2} & 0 & 0 & \frac{1-q}{2} \\
0 & q & 0 & 0 \\
0 & 0 & q & 0 \\
\frac{1-q}{2} & 0 & 0 & \frac{1+q}{2}
\end{pmatrix},\quad q=(-\frac{1}{3})^{N-1}.
\end{eqnarray}
Let us generalize the random bond to an arbitrary $SU(2)$ rotation:
\begin{eqnarray}
V=\begin{pmatrix}
\alpha & \beta \\[6pt]
-\beta^{*} & \alpha^{*}
\end{pmatrix},\quad |\alpha|^2 + |\beta|^2 = 1.
\end{eqnarray}
Applying random bond between site $N$ and $1$, we get  $M^{(N)} (U \otimes U^*) $ , which is:
\begin{eqnarray}
&&M^{(N)} (U \otimes U^*)=\\
&&
\begin{pmatrix}
\frac{1+q}{2}-q|\beta|^{2} & q\,\alpha\beta^{*} & q\,\beta\alpha^{*} & \frac{1-q}{2}+q|\beta|^{2} \\
-q\,\alpha\beta & q\,\alpha^{2} & -q\,\beta^{2} & q\,\alpha\beta \\
-q\,\alpha^{*}\beta^{*} & -q\,(\beta^{*})^{2} & q\,(\alpha^{*})^{2} & q\,\alpha^{*}\beta^{*} \\
\frac{1-q}{2}+q|\beta|^{2} & -q\,\alpha\beta^{*} & -q\,\beta\alpha^{*} & \frac{1+q}{2}-q|\beta|^{2}
\end{pmatrix}. \nonumber
\end{eqnarray}
The density matrix on site $1$ is:
\begin{eqnarray}
&&\text{tr}_{2\dots N}U\rho_{\text{AKLT},V}U^{\dagger} \\
&&=\sum_{s,s'=0,\pm1}|s\rangle \langle s'|\text{tr}( A_s  \otimes A^*_{s'} )M^{(N)} (V \otimes V^*)\propto\nonumber\\
&&
\!\!\!\!\!\!\!\!\!\begin{pmatrix}
\frac{1-q}{2}+q|\beta|^{2}
& i\sqrt{2}\,q\,\beta\,\operatorname{Im}(\alpha)
& q\,\beta^{2}
\\[4pt]
-i\sqrt{2}\,q\,\beta^{*}\operatorname{Im}(\alpha)
& \frac{1+q}{2}-q|\beta|^{2}-q\,\operatorname{Re}(\alpha^{2})
& -i\sqrt{2}\,q\,\beta\,\operatorname{Im}(\alpha)
\\[4pt]
q\,(\beta^{*})^{2}
& i\sqrt{2}\,q\,\beta^{*}\operatorname{Im}(\alpha)
& \frac{1-q}{2}+q|\beta|^{2}
\end{pmatrix}.\nonumber
\end{eqnarray}
Compared to the case when $V=\sigma_I$:
\begin{eqnarray}
\!\!\!\!\!\!\!\!\!\!\!\!\!\text{tr}_{2\dots N}(\rho_{\text{AKLT}})&=&\sum_{s,s'=0,\pm1}|s\rangle \langle s'|\text{tr}( A_s  \otimes A^*_{s'} )M^{(N)} \\
&\propto&
\begin{pmatrix}
\frac{1-q}{2} & 0 & 0 \\
0 & \frac{1-q}{2} & 0 \\
0 & 0 & \frac{1-q}{2}
\end{pmatrix},
\end{eqnarray}
which gives 
\begin{eqnarray}
&&U\rho_{\text{AKLT},V}U^{\dagger} = \rho_{\text{AKLT}}, \\
\implies&&
    \alpha=\pm1,\;\beta=0,\\
    \iff&& V = \pm \sigma_I.
\end{eqnarray}
Thus, we showed that for any nontrivial random bond $V\neq \pm \sigma_I$, there does not exist any semi-local unitary $U$ that can reduce the random bond AKLT to the original AKLT state.

\section{Proof that the post-POVM states are encoded graph states}
\label{app:oldproof}

We discuss in the main text that random bounds arise from measuring the two dangling qubits from nearby blocks in the Bell-state basis. In that part, we would like to obtain the outcome of a singlet so that the two blocks are connected. But because of other outcomes, the bond can be a triplet. Thus, it is natural to consider the random bond. Furthermore, the random decoration of a spin-1 decoration on an edge can be reduced to a random bond. Thus, we can just consider the generalization of the original AKLT construction to use randomly any of the four Bell states (singlet and triplet entangled states) and discuss the computational capability of the resultant random-bond AKLT state, e.g., on the hexagonal lattice.  Below, we prove that the post-POVM states for the random-bond AKLT state are still an encoded random graph state. Rigorously proving computational universality may require sampling random POVM outcomes and then performing percolation simulations on the resulting random graphs. When sampling the POVM, we first randomly assign the bonds, use Metropolis sampling, and need to reject cases where a loop has frustration; otherwise, the rest is similar to the previous simulations~\cite{WeiAffleckRaussendorf11,Wei13}. Moreover, we emphasize that our proof directly applies to show that, in general,
\begin{equation}
|\Psi_G\rangle\equiv\mathop{\bigotimes}_v F_{\alpha_v,v}\mathop{\bigotimes}_e|{\rm Bell\, state}\rangle_e
\end{equation}
is an encoded graph state, going beyond the case of spin-3/2, as the form of $F$ for spin-$S$ can be easily incorporated, where we have suppressed the projection $P_v$ (\ref{eq:Projection}) in the above equation, as $F_{\alpha_v,v} P_v=F_{\alpha_v,v}$.

We remark that the proof for the random-bond case that the post-POVM state is an encoded graph state is almost identical to the original proof for the uniform singlet-bond case; see Ref.~\cite{wei2012two}. \textit{Our presentation here will mimic the structure and the notation there in most parts.} But we shall supply with more explanation whenever needed.

Let us first recall that the POVM to be performed on all sites can be expressed in terms of the virtual qubit degrees of freedom:
\begin{subequations}
\label{POVM2}
  \begin{eqnarray*}
{F}_{v,z}\!\!\!&=&\!\!\sqrt{\frac{2}{3}}\Big(\ketbra{000}+\ketbra{111}\Big) \\
{F}_{v,x}\!\!\!&=&\!\!\sqrt{\frac{2}{3}}\Big(\ketbra{\!+\!+\!+\!}+\ketbra{\!-\!-\!-\!}\Big)\\
{F}_{v,y}\!\!\!&=&\!\!\sqrt{\frac{2}{3}}\Big(\ketbra{i,i,i}+\ketbra{\!\!-\!i,\!-i,-i}\Big),
\end{eqnarray*}
\end{subequations}
which can be rewritten as:
\begin{equation}
  \label{POVM2a}
  \begin{array}{rcl}
    F_{v,z} &=& \displaystyle{\sqrt{\frac{2}{3}}\frac{I_{12}+Z_1Z_2}{2}\frac{I_{23}+Z_2Z_3}{2}},\\
    F_{v,x} &=& \displaystyle{\sqrt{\frac{2}{3}}\frac{I_{12}+X_1X_2}{2}\frac{I_{23}+X_2X_3}{2}},\\
    F_{v,y} &=&
    \displaystyle{\sqrt{\frac{2}{3}}\frac{I_{12}+Y_1Y_2}{2}\frac{I_{23}+Y_2Y_3}{2}},
  \end{array}
\end{equation}
where the indices 1,2, and 3 indicates which virtual qubit.
When expanded, the above $F$'s is expressed in terms of the sum of the stabilizer group ${\cal S}_F=\{I, \sigma^{(1)}\sigma^{(2)}, \sigma^{(1)}\sigma^{(3)}, \sigma^{(2)}\sigma^{(3)}\}$, where $\sigma$ is one of the three Pauli matrices, depending on the POVM outcome. In order to demonstrate that an operator is a stabilizer operator for the post-POVM state, that operator needs to commute with the corresponding POVM operator $F$ on all sites, i.e., the corresponding stabilizer group ${\cal S}_F$.

The $F$ for arbitrary spin-$S$ (i.e., $2S$ virtual qubits) can be constructed as follows.
\begin{equation}
F_\alpha^{(S)} \sim \prod_{i=1}^{2S-1} (I+ \sigma_\alpha^{(i)}\sigma_\alpha^{(i+1)})/2.
\end{equation}
Namely, it is constructed with $2S-1$ independent stabilizer operators $\sigma_\alpha^{(i)}\sigma_\alpha^{(i+1)}$.

It turns out that, for any POVM outcome labeled by ${\cal{A}}=\{\alpha_i={\rm x,y, or\, z}\}$, the post-POVM state
$|\Psi({\cal{A}})\rangle=\otimes_i F_{i,\alpha_i} \otimes_e|{\rm Bell\, state}\rangle_e$ is locally equivalent to an encoded graph
state $\overline{|G({\cal{A}})\rangle}$, where $i$ is the index of a site and $\alpha_i$ labels the corresponding POVM outcome. The Bell states in the product $\otimes_e|{\rm Bell\, state}\rangle_e$ need not be the same one and can be chosen randomly (hence, the adjective random-bond), as we consider here.  The graph
$G({\cal{A}})$ can be constructed as follows.
First, we refer to an edge $(v,w) \in
E({\cal{L}})$, the edge set of lattice ${\cal L}$, as an internal iff at the sites $v$ and $w$ the
local POVM has resulted in the same outcome $\alpha$. The graph
$G({\cal{A}})$ is constructed by modifying the lattice graph ${\cal{L}}$ by first (1)
contracting all internal edges, and, in the resultant multi-graph,
(2a) deleting all edges of even multiplicity and (2b) converting all
edges of odd multiplicity into the conventional edges of multiplicity 1; see Fig.~\ref{fig:AKLToutcomesGraph} for illustration.

\begin{figure}
\centering
\includegraphics[width=0.45\textwidth]{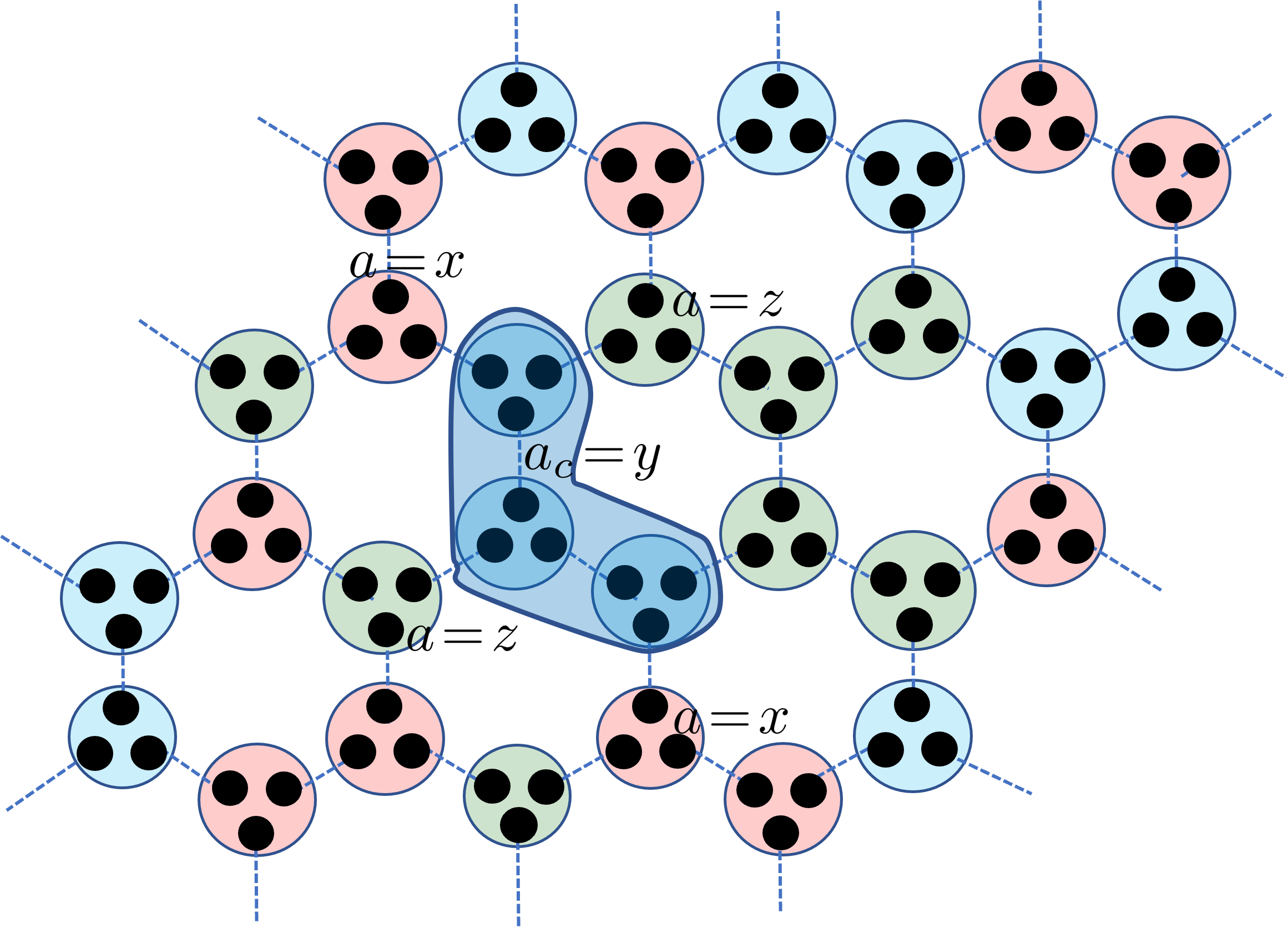}

\caption{Illustration of POVM outcomes and a particular domain with $a_c=y$. } 
\label{fig:POVMdomains}
\end{figure}
In step (1) of the above procedure, several connected sites of ${\cal{L}}$ with the same POVM outcome are
merged into a single composite object ${\cal{C}} \in
V(G({\cal{A}}))$, which we also refer to as a \textit{domain}. Each such domain ${\cal{C}}$ is a {\em{vertex}} in
the graph $G({\cal{A}})$.  Physically, in a domain of type
$a$, we have a mixture of ferromagnetic and antiferromagnetic order along the $\pm a$-direction, and whether the neighboring spins align or anti-align depends on the bond between them; see Table~\ref{tbl:BellStates}, where the sign in front of the stabilizer $\sigma\otimes\sigma$ gives whether  
because two neighboring spins align (+) or anti-align (-).  The state of the domain
contains only two complimentary configurations w.r.t. the quantization axis $a$, e.g.,
$|+3/2,+3/2,-3/2,\dots\rangle$ and $|-3/2,-3/2,+3/2,\dots\rangle$, which constitute one qubit effectively .

As shown in Refs.~\cite{WeiAffleckRaussendorf11,wei2012two},  the computational power of the AKLT state hinges on the connectivity properties
of $G({\cal{A}})$: if the graphs are percolated, then we can further perform local measurements to distill a regular cluster state, which is universal for measurement-based quantum computation.  This implies that the original AKLT state is universal. However, if the resultant graphs after the POVM are not percolated, the resultant graph states are not universal. But one cannot directly conclude that the original AKLT state is not universal, because there might exist other measurement schemes to yield computational power. However, this was never explicitly constructed.

We now follow the original three-step proof in Ref.~\cite{wei2012two}. As we argued previously, each domain ${\cal{C}} \in V(G({\cal{A}}))$ gives rise to one encoded qubit, and we give a proof of this.

{\em{Step 1: Encoding.}} Consider a domain ${\cal{C}} \subset
V({\cal{L}})$. That is, Since all sites $v$ in domain  ${\cal{C}}$ has the same POVM
outcome $a \in \{x,y,z\}$, and this means that there are $2|{\cal{C}}|$ (independent) stabilizer generators, c.f.
Eq.~(\ref{POVM2}), enforced by the projections $F_{v,a}$ on all $v \in
{\cal{C}}$. Moreover, we can choose a tree ${\cal{T}}$ (no loops) among the
set of edges with both endpoints in the domain ${\cal{C}}$ and  each
edge $(u,v) \in {\cal{T}}$ contributes a stabilizer generator
$\pm\sigma_a^{(u)}\sigma_a^{(v)}$ to the product of Bell states; see Table~\ref{tbl:BellStates}. There are $|{\cal{T}}| = |{\cal{C}}|-1$ such constraints. Thus, in total (together with those from the POVM), there are $3|{\cal{C}}|-1$
stabilizer generators with support only in ${\cal{C}}$, acting on
$3|{\cal{C}}|$ qubits,   giving rise to {\it one encoded qubit\/}. (We note that in the case of spin-$S$, there are $(2S-1)|{\cal C}|$ independent stabilizer generators from the projections $F$'s on $v \in {\cal C}$ and there are $|{\cal C}|-1$ independent stabilizer generators from the internal links. Hence, there is one encoded qubit in a domain ${\cal C}$. It generalizes to the case where each site may have a different spin magnitude $S$.) 

While the stabilizer generators for our code follow from the
construction, there is freedom in choosing the encoded Pauli
operators. Table~\ref{tbl:coding2} shows one such choice of
encoding.

In the following second step, we show that $|\Psi({\cal{A}})\rangle$
is, up to local encoded unitaries, equivalent to the encoded graph
state $\overline{|G({\cal{A}})\rangle}$.

{\em{Step 2:}} Here, we adopt the same definitions as in Ref.~\cite{wei2012two}. Consider a central vertex ${\cal{C}}_c \in
V(G({\cal{A}}))$ and all its neighboring vertices ${\cal{C}}_\mu \in
V(G({\cal{A}}))$. Denote the POVM outcome for all ${\cal{L}}$-sites
$v \in {\cal{C}}_c, {\cal{C}}_\mu$ by $a_c$ and $a_\mu$,
respectively; see Fig.~\ref{fig:POVMdomains} for illustration. Denote by $E_\mu$ the set of ${\cal{L}}$-edges that
run between ${\cal{C}}_c$ and ${\cal{C}}_\mu$. Denote by $E_c$ the
set of ${\cal{L}}$-edges internal to ${\cal{C}}_c$. Denote by $C_c$
the set of all qubits in ${\cal{C}}_c$, and by $C_\mu$ the set of
all qubits in ${\cal{C}}_\mu$. 

We first consider the stabilizer of the state $\bigotimes_{e \in
E({\cal{L}})}|{\rm Bell\, state}\rangle_e$. For any $\mu$ and any edge $e \in
E_\mu$, let $u(e)$ [$v(e)$] be the endpoint of $e$ in $C_\mu$
[$C_c$]. Then, for all $\mu$ and all $e \in E_\mu$ the Pauli
operators $\pm\sigma_{a_\mu}^{({u(e)})}\sigma_{a_\mu}^{(v(e))}$ are in
the stabilizer of $\bigotimes_{e \in E({\cal{L}})}|{\rm Bell\, state}\rangle_e$, where the sign depends on the bond or the corresponding Bell state; see Table~\ref{tbl:BellStates}. Now, we
choose $b \in \{x,y,z\}$ such that $b \neq a_c$, and let, for any
edge $e' \in E_c$, $v_1(e'), v_2(e') \in C_c$ be qubit locations
such that $e'=(v_1(e'), v_2(e'))$. 
(Note that since there are two possible choices, we can use the freedom to fix the choice so that the local operator of the final encoded stabilizer operator associated with the domain is $\pm \overline{X}$ when restricting the support to this domain.)
Then, for all $e' \in E_c$,
$\pm\sigma_b^{({v_1(e')})}\sigma_b^{(v_2(e'))}$  is in the stabilizer
of $\bigotimes_{e \in E({\cal{L}})}|{\rm Bell\, state}\rangle_e$, where, again, the sign needs to be chosen according to the underlying bond. Therefore, the
product of all these operators,
\begin{equation}
  O_{{\cal{C}}_c} = \pm \! \left(\bigotimes_{\mu}\bigotimes_{e \in E_\mu} \sigma_{a_\mu}^{(u(e))} \sigma_{a_\mu}^{(v(e))} \! \right)\!\! \left(\bigotimes_{e' \in E_c} \sigma_{b}^{(v_1(e'))} \sigma_{b}^{(v_2(e'))} \!\!\right)
\end{equation}
is also in the stabilizer of $\bigotimes_{e \in
E({\cal{L}})}|{\rm Bell\, state}\rangle_e$.

We now show that $O_{{\cal{C}}_c}$ commutes with the local POVMs and
is therefore also in the stabilizer of $|\Psi({\cal{A}})\rangle$. The argument here is exactly identical to that in Ref.~\cite{wei2012two}, so we reproduce it here without the need of modification.
First, consider the central domain ${\cal{C}}_c$. The operator
$O_{{\cal{C}}_c}$ acts non-trivially on every qubit in $C_c$,
$O_{{\cal{C}}_c}|_l \neq I_l$ for all qubits $l \in C_c$.
Furthermore, for all qubits $l \in C_c$,
$O_{{\cal{C}}_c}\left|_l\right. \neq \sigma_{a_c}^{(l)}$. Namely, if
$l \in C_c$ is connected by an edge $e \in E_\mu$ to
${\cal{C}}_\mu$, for some $\mu$, then
$O_{{\cal{C}}_c}\left|_l\right. = \sigma_{a_\mu}^{(l)}  \neq
\sigma_{a_c}^{(l)}$ (for all $\mu$, $a_\mu \neq a_c$ by construction
of $G({\cal{A}})$). Or, if $l \in C_c$ is the endpoint of an
internal edge $e' \in E_c$ then $O_{{\cal{C}}_c}\left|_l\right. =
\sigma_b^{(l)}  \neq \sigma_{a_c}^{(l)}$ ($a_c \neq b$ by above
choice). Therefore, for any $i,j \in C_c$, $O_{{\cal{C}}_c}$
anticommutes with $\sigma_{a_c}^{(i)}$ and $\sigma_{a_c}^{(j)}$, and
thus commutes with all $\sigma_{a_c}^{(i)}\sigma_{a_c}^{(j)}$. Thus,
$O_{{\cal{C}}_c}$ commutes with the local POVMs Eq.~(\ref{POVM2a})
on all $v \in {\cal{C}}_c$.

Next, consider the neighboring  domains ${\cal{C}}_\mu$.
$O_{{\cal{C}}_c}\left|_{{\cal{C}}_\mu}\right. =
\bigotimes_{j}\sigma_{a_\mu}^{(j)}$ by construction.
$O_{{\cal{C}}_c}$ thus commutes with the local POVMs $F_{v,a_\mu}$
for all $v \in {\cal{C}}_\mu$ and for all $\mu$. 

Therefore, $O_{{\cal{C}}_c}$ is in the  stabilizer of
$|\Psi({\cal{A}})\rangle$. Therefore, $O_{{\cal{C}}_c}$ is an
encoded operator w.r.t. the code in Table~\ref{tbl:coding2}, and we
need to figure out which form of the logical X that we will take. (1) Central vertex ${\cal{C}}_c$:
$\left. O_{{\cal{C}}_c}|_{C_c}\right.$ is an encoded operator on
$C_c$, $\left. O_{{\cal{C}}_c}|_{C_c}\right. \in \{\pm \overline{I},
\pm \overline{X}, \pm \overline{Y}, \pm \overline{Z}\}$. Since
$\left. O_{{\cal{C}}_c}|_l\right. \neq \sigma_{a_c}^{(l)}$ for any
$l \in {\cal{C}}_c$, by Table~\ref{tbl:coding2}, $\left.
O_{{\cal{C}}_c}|_{C_c}\right. \neq \pm \overline{I}, \pm
\overline{Z}$. Thus, $\left. O_{{\cal{C}}_c}|_{C_c}\right. \in \{\pm
\overline{X}, \pm \overline{Y}\}$. 
Different from the original proof in Ref.~\cite{wei2012two}, here we will take advantage of the freedom to choose $b$ so that $O_{{\cal{C}}_c}|_{C_c}=\pm
\overline{X}$. If the outcome were $\pm\overline{Y}$, we could change the local $b$ to the other alternative, and this change only affects the local logical operator at this domain. (2) For the neighboring vertices
${\cal{C}}_\mu$,  by using  Table~\ref{tbl:coding2},  we see that $\sigma_{a_\mu}^{(l)} =
\pm \overline{Z}$, for any $l \in C_\mu$. Thus, $\left.
O_{{\cal{C}}_c}|_{C_\mu}\right. = \pm \overline{Z}^{|E_\mu|}$. Now
observe that $Z^2 = I$, and that, this justifies the above
prescription in constructing the graph $G({\cal{A}})$. Using the
adjacency matrix $A_{G({\cal{A}})}$, we have $|E_\mu| \mod 2 =
[A_{G({\cal{A}})}]_{c,\mu}$ and hence $\left.
O_{{\cal{C}}_c}|_{C_\mu}\right. =\pm
\overline{Z}^{[A_{G({\cal{A}})}]_{c,\mu}}$. This justifies the modulo-2 rule for edges between two domains.

Thus, using the freedom in Table~\ref{tbl:coding2}, we have proved that, for all ${\cal{C}}_c \in V(G({\cal{A}}))$,
\begin{equation}
  \label{CSstab}
  O_{{\cal{C}}_c} \in \left\{ \pm \overline{X}_{C_c} \bigotimes_{{\cal{C}}_\mu \in V(G({\cal{A}}))}  \overline{Z}_{{\cal{C}}_\mu}^{[A_{G({\cal{A}})}]_{c,\mu}}\right\}.
\end{equation}
The code stabilizers in
Table~\ref{tbl:coding2}
 and the stabilizer operators in Eq.~(\ref{CSstab})
together define the state $|\Psi({\cal{A}})\rangle$ uniquely.
Thus, $|\Psi({\cal{A}})\rangle$ is an encoded graph state $|G({\cal{A}})\rangle$.

Finally, we show that the
encoding can be unraveled by local spin-3/2 measurements.

{\em{Step 3: Decoding of the code.}}  We now show that any domain
${\cal{C}} \in V(G({\cal{A}}))$ can be reduced to a single
elementary site $w \in V({\cal{L}})$ by local measurement on all
other sites $v \in {\cal{C}}$, $v \neq w$. This means that subsequent logical operation after the reduction can be performed on single sites. For any such $v$, choose
the measurement basis ${\cal{B}}_a$, $a \in \{x,y,z\}$, as follows
\begin{equation}
  \label{Meas}
  \begin{array}{rcl}
    {\cal{B}}_x &=&\left\{{(|+++\rangle \pm |---\rangle)}/{\sqrt{2}} \right\}, \\
    {\cal{B}}_y &=& \left\{{(|i,i,i\rangle \pm|-i,-i,-i\rangle)}/{\sqrt{2}}\right\}, \\
    {\cal{B}}_z&=&\left\{{(|000\rangle \pm |111\rangle)}/{\sqrt{2}}\right\}.
  \end{array}
\end{equation}
These measurements map the symmetric subspace of the three-qubit
states into itself, and  they can therefore be performed on the
physical spin 3/2 systems, up to a correction of the logical Z operator. (This can be straightforwardly extended to spin $S$.) In a more plain language, according to its POVM outcome label $a$ of the domain with two sites, e.g., $a=z$, an encoded logical state may be $\alpha|000\rangle|111\rangle +\beta|111\rangle |000\rangle$. One measures in $B_z$ above, e.g., to remove the second part and reduces the encoded state to $\alpha|000\rangle\pm \beta|111\rangle$.

Denote by ${\cal{S}}_{\cal{C}}$ and ${\cal{S}}_{{\cal{C}}\backslash
v}$ the code stabilizer on the domain ${\cal{C}} \in
V(G({\cal{A}}))$ and on the reduced domain ${\cal{C}}\backslash v$,
respectively. Using standard stabilizer techniques, the measurement Eq.~(\ref{Meas})  has the
following effect on the encoding
\begin{equation}
    {\cal{S}}_{\cal{C}} \longrightarrow  {\cal{S}}_{{\cal{C}}\backslash v},\;
    \overline{X}_{\cal{C}} \longrightarrow \pm \overline{X}_{{\cal{C}}\backslash v},\;
    \overline{Z}_{\cal{C}} \longrightarrow \overline{Z}_{{\cal{C}}\backslash v}.
\end{equation}
The measurement (\ref{Meas}) thus removes from ${\cal{C}}$ by  one
lattice site $v \in V({\cal{L}})$. We repeat the procedure until
only one site, $w$, remains in ${\cal{C}}$, for each ${\cal{C}} \in
V(G({\cal{A}}))$. In this way, ${\cal{S}}_{\cal{C}} \longrightarrow
{\cal{S}}_{\{w\}}$, $\overline{X}_{\cal{C}} \longrightarrow \pm
\overline{X}_{\{w\}}$, $\overline{Z}_{\cal{C}}  \longrightarrow
\overline{Z}_{\{w\}}$. Thus, $|\Psi(A)\rangle \longrightarrow
\overline{U}_\text{loc} \overline{|G(A)\rangle}=: |G(A)\rangle$,
where $U_\text{loc}$ is a local unitary, and the encoding in
Table~\ref{tbl:coding2} has now shrunk to one site of ${\cal{L}}$
per encoded qubit, i.e., to three auxiliary qubits. This part is also identical to that in the original proof in Ref.~\cite{wei2012two}. 

To perform quantum computation on the remaining encoded qubits, one can identify paths that represent the evolution of logical qubits under discrete single-logical-qubit gates. These paths need to be sufficiently separated from each other, and there should be multiple segments connecting neighboring paths.  In another way, one can reduce the random graph state to a regular square lattice cluster state, which was proved to be possible if the graph resides in the percolation phase~\cite{wei2012two}. Subsequently, one just follows the procedure in the one-way quantum computation using the cluster state~\cite{raussendorf2001a}. Note that the measurement of an encoded qubit on
a site $w \in {\cal{L}}$ is an operation on the symmetric subspace
of three auxiliary qubits at $w$, and can thus be realized as a
measurement of the equivalent physical spin. This applies to arbitrary spin-$S$. However, whether for $S>2$, the measurement elements involving just the $F$'s (or related combination) may not be sufficient to complete the POVM, which means that there may be leakage outside the logical qubit space (see Refs.~\cite{WeiPoyaRaussendorf,WeiRaussendorf15} the treatment for spin-2 case), which is left for future exploration.

\bibliography{references}

\end{document}